%% file: main.tex
\documentclass[journal]{IEEEtran}
\IEEEoverridecommandlockouts
\usepackage{cite}
\usepackage[colorlinks=true,linkcolor=black,anchorcolor=black,citecolor=black,filecolor=black,menucolor=black,runcolor=black,urlcolor=magenta]{hyperref}

\usepackage{amsmath,amssymb,amsfonts,mathrsfs}
\usepackage{algorithmic}
\usepackage{graphicx}
\usepackage{textcomp}
\usepackage{xcolor}
\usepackage{booktabs}
\usepackage{subeqnarray}
\usepackage{multicol}
\def\BibTeX{{\rm B\kern-.05em{\sc i\kern-.025em b}\kern-.08em
		T\kern-.1667em\lower.7ex\hbox{E}\kern-.125emX}}

\ifCLASSOPTIONcompsoc
\usepackage[caption=false,font=normalsize,labelfont=sf,textfont=sf]{subfig}
\else
\usepackage[caption=false,font=footnotesize]{subfig}
\fi

\newcommand{\RE}[1]{\operatorname{Re}\left({#1}\right)}
\newcommand{\IM}[1]{\operatorname{Im}\left({#1}\right)}

\newcommand{\mat}[1]{\boldsymbol{#1}}
\newcommand{\bs}[1]{\boldsymbol{#1}}

\begin{document}
	\bstctlcite{Settings}
\title{Validation of the Reference Impedance in Multiline Calibration with Stepped Impedance Standards}

\author{%
	\IEEEauthorblockN{%
		Ziad~Hatab$^1$, Michael~Ernst~Gadringer$^1$, Ahmad~Bader~Alothman~Alterkawi$^2$, and~Wolfgang~Bösch$^1$\\
		$^1$Graz University of Technology, Graz, Austria\\
		$^2$AT\&S, Leoben, Austria\\
		\{z.hatab; michael.gadringer; wbosch\}@tugraz.at, \ a.alterkawi@ats.net
	}%
\thanks{Software implementation and measurements are available online:\\ \url{https://github.com/ZiadHatab/verification-multiline-trl-calibration}}
}%
\markboth{This work has been accepted for publication in the IEEE Open Journal of Instrumentation and Measurement}{}
\maketitle

\begin{abstract}
This paper presents a new technique for evaluating the consistency of the reference impedance in multiline thru-reflect-line (TRL) calibration. During the calibration process, it is assumed that all transmission line standards have the same characteristic impedance. However, these assumptions are prone to errors due to imperfections, which can affect the validity of the reference impedance after calibration. Our proposed method involves using multiple stepped impedance lines of different lengths to extract the broadband reflection coefficient of the impedance transition. This reflection coefficient can be used to validate the reference impedance experimentally without requiring fully defined standards. We demonstrate this method using multiline TRL based on microstrip structures on a printed circuit board (PCB) with an on-wafer probing setup.
\end{abstract}


\input{Sections/Section1}
\input{Sections/Section2}
\input{Sections/Section3}
\input{Sections/Section4}
\input{Sections/Section5}

\input{Sections/Section6}


\section*{Acknowledgment}
The financial support by the Austrian Federal Ministry for Digital and Economic Affairs and the National Foundation for Research, Technology, and Development is gratefully acknowledged. The authors thank AT\&S for manufacturing the PCBs and providing cross-sectional images of the microstrip lines, and ebsCENTER for granting access to their measurement equipment.

\bibliographystyle{IEEEtran}
\bibliography{References/references.bib}

\end{document}

%% file: Sections/Section1.tex
\section{Introduction}
\label{sec:1}
\IEEEPARstart{T}{he} accuracy of S-parameter measurements performed by a vector network analyzer (VNA) heavily depends on the calibration method and standards used. The multiline thru-reflect-line (TRL) calibration procedure, first introduced by the National Institute of Standards and Technology (NIST) \cite{Marks1991}, provides a precise definition of the calibration plane. The high accuracy of multiline TRL method comes from the fact that transmission line standards can be manufactured with high precision, even with conventional machinery capabilities, in contrast to resistive load standards. 

Experimental validation of the calibration is a crucial aspect of any calibration technique. The multiline TRL method is a self-calibration technique, where some of the calibration standards are not fully specified in advance. However, the multiline TRL algorithm requires consistency among the line standards. This means that the line standards should be identical in all aspects except for their length. Inconsistency between line standards can lead to impedance mismatch \cite{Ye2017,Hyunji2021,Lenk2013}.

Traditional calibration validation techniques require complete knowledge of a set of reference standards. These techniques are often used in coaxial or waveguide interfaces where traceable standards are available \cite{Zeier2018,Ridler2019,Ridler2021}. However, multiline TRL calibration is also used for planar circuits, such as printed circuit boards (PCBs) and wafers, where the fabrication of traceable standards is more challenging. Commercial impedance standard substrates (ISSs) are commonly used for on-wafer calibration. These standards can be characterized using traceable on-wafer standards, as demonstrated in \cite{Arz2019,Arz2019a}. On the other hand, establishing traceable standards with PCB technology is more challenging. PCB manufacturing mostly uses composite materials made from reinforced fiberglass with epoxy resin, which can introduce significant delay skew depending on the location of the transmission lines on the substrate \cite{Shlepnev2014,Chen2019}. Manufacturing tolerances of PCBs are also much higher than those in on-wafer applications \cite{Lau2019,Sepaintner2020}. 

\begin{figure}[th!]
	\centering
	\subfloat[]{\includegraphics[width=.48\linewidth]{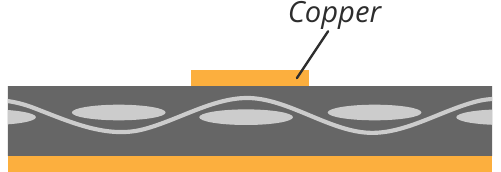}\label{fig:1.1a}}~
	\subfloat[]{\includegraphics[width=.48\linewidth]{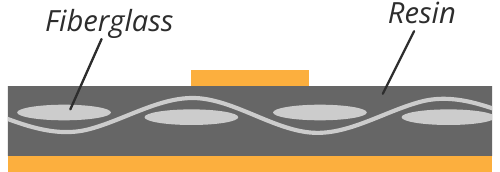}\label{fig:1.1b}}
    \caption{An illustration of a cross-section of a microstrip on a PCB with a substrate based on reinforced fiberglass in epoxy resin. (a) The microstrip line is placed directly above the fiberglass. (b) The microstrip line is placed between the glass yarns.}
	\label{fig:1.1}
\end{figure}

An illustration of microstrip lines on a PCB based on a mixture of fiberglass and epoxy resin is shown in Fig.~\ref{fig:1.1}. As part of our experimental measurements, we analyzed the cross-section of microstrip lines on a Megtron 7 substrate, as shown in Fig.~\ref{fig:5.2}. Previous studies have also presented cross-sectional images of fiberglass dielectric substrates \cite{Chen2019,Manukovsky2019}.

The calibration comparison method \cite{Williams1991, Rumiantsev2006} is a widely used approach for validating on-wafer multiline TRL calibration. This method involves comparing the multiline TRL calibration to another multiline TRL calibration on a different reference substrate. A major drawback of this method is the requirement of a fully characterized calibration kit. Furthermore, the method is sensitive to parasitic inductance resulting from pads and material differences between the calibration kits, as demonstrated in \cite{Williams2001, Galatro2017}.

A common technique for verifying multiline TRL calibration is to investigate the calibrated reflect standard \cite{EURAMET2018}. During calibration, the reflect standard is not specified but identical at both ports. Ideally, the calibrated reflect standard should exhibit behavior similar to that of an ideal reflect standard (e.g., short or open). However, using the calibrated reflect standard as a validation metric has a couple of shortcomings. First, predicting the parasitic behavior of the reflect standard at high frequencies can be challenging, especially at millimeter-wave and beyond. Second, even if the reflect standard exhibits an ideal response, this does not inform us about the accuracy of the reference impedance. Ideal short or open standards are impedance independent. While a reflect standard might not be advantageous for validating the reference impedance, it can be useful in identifying user error during calibration.

Another commonly used two-port device for calibration validation is the stepped impedance line (also known as the Beatty line) \cite{EURAMET2018}. This device is widely used in the validation of airline coaxial multiline TRL calibration. The problem with using a single stepped impedance line is that its desired mismatch behavior is bandwidth-limited due to its physical length. In fact, maximum mismatch occurs at integer multiples of quarter-wavelengths. Using a shorter stepped line can increase the mismatch bandwidth, but at the cost of the uncertainty at lower frequencies. Additionally, the abrupt change in the impedance introduces some parasitic effects that are more apparent at higher frequencies, which could lead to a false interpretation of the mismatch.

In this paper, we adapt the method of using the stepped impedance line for validation by increasing the mismatch bandwidth and excluding the parasitic effect of the abrupt impedance change. The bandwidth limitation of the mismatch response due to the length of the line is no different from that of TRL calibration. Hence, we propose using a second multiline TRL calibration using a set of stepped impedance lines of different lengths. By using both calibrations with matched and mismatched lines, we can extract the broadband reflection coefficient of the impedance transition. Additionally, we introduce a modeling approach for the parasitic behavior of the impedance jump to eliminate its impact on the extracted reflection coefficient.

The goal of extracting the broadband reflection coefficient of the impedance transition is to use it as a validation metric. This has a couple of advantages. First, the extracted reflection coefficient is due to a finite impedance mismatch (not open or short), meaning that the results of the reflection coefficient depend on the reference impedance of the multiline TRL that we want to validate. Therefore, any variation in the reference impedance of the first multiline TRL should clearly translate to the extracted reflection of the impedance transition. Secondly, since we are extracting a broadband response of the mismatch of the stepped line standards, the frequency response of the reflection coefficient has a near-flat frequency response, even for quasi-TEM (transverse electromagnetic) structures as microstrip lines, as will be demonstrated in Sections~\ref{sec:3} and \ref{sec:5}. Finally, for PCB applications, the stepped lines are implemented on the same substrate as the matched lines. Hence, we don't require any prior characterization of the standards. We can think of the proposed method as artificially measuring a broadband mismatch load. The validation of the calibration is accomplished by defining validation bounds through uncertainty propagation in the geometrical and material variation of the cross-section of the transmission lines.

The remainder of this paper is structured as follows. Section~\ref{sec:2} discusses the residual error in calibration resulting from impedance variation in the calibration standards. Next, in Section~\ref{sec:3}, we present the mathematical derivation of the proposed validation method. We then introduce the validation bounds in Section~\ref{sec:4}. Finally, Sections~\ref{sec:5} and \ref{sec:6} discuss PCB measurements and provide concluding remarks.


%% file: Sections/Section2.tex
\section{Analysis of Impedance Mismatch on Calibration}
\label{sec:2}

\begin{figure}[th!]
	\centering
	\includegraphics[width=.95\linewidth]{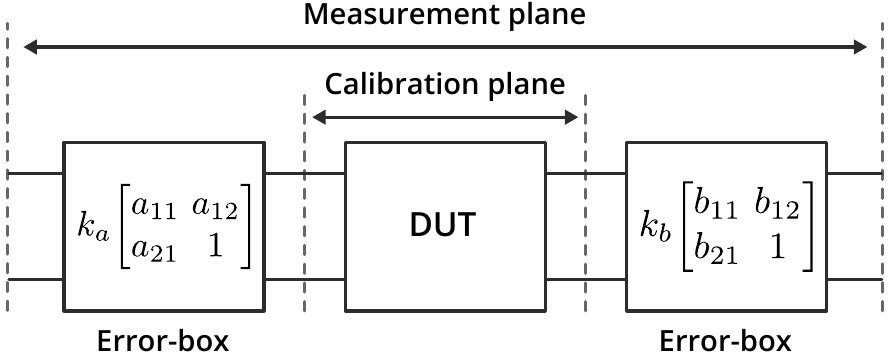}
	\caption{Illustration of the error box model of a two-port VNA.}
	\label{fig:2.1}
\end{figure}

The purpose of this section is to highlight the impact of mismatch on calibration, and to identify the appropriate standards to be used for validation purposes. We begin the analysis by defining the error box model of a two-port VNA, as shown in Fig.~\ref{fig:2.1}. The measurement of a device under test (DUT) with an uncalibrated VNA can be expressed using T-parameters as follows:
\begin{equation}
	\mat{M}_\mathrm{dut} = \underbrace{k_ak_b}_{k}\underbrace{\begin{bmatrix} a_{11} & a_{12}\\a_{21} & 1\end{bmatrix}}_{\mat{A}}\mat{T}_\mathrm{dut}\underbrace{\begin{bmatrix} b_{11} & b_{12}\\b_{21} & 1\end{bmatrix}}_{\mat{B}},
	\label{eq:2.1}
\end{equation}
where the matrices $\mat{A}$ and $\mat{B}$ represent the T-parameters of the individual port error boxes holding the first six error terms, and $k$ represents the 7th error term describing the transmission error between the ports.

After performing a multiline TRL calibration, we obtain estimates for the calibration coefficients $\mat{A}$, $\mat{B}$, and $k$. However, since measurements are never perfect, the estimates of the calibration coefficients are prone to errors, which can be described by the following notation:
\begin{equation}
    \widehat{\mat{A}} = \mat{A}\widetilde{\mat{A}}^{-1}\frac{1}{\widetilde{k}_a}, \qquad \widehat{\mat{B}} = \frac{1}{\widetilde{k}_b}\widetilde{\mat{B}}^{-1}\mat{B},
    \label{eq:2.2}
\end{equation}

Here, the matrices $\widehat{\mat{A}}$ and $\widehat{\mat{B}}$ represent the estimated error boxes obtained from the calibration process. On the other hand, the matrices $\widetilde{\mat{A}}$ and $\widetilde{\mat{B}}$ summarize the residual error and are defined as follows:
\begin{equation}
    \widetilde{\mat{A}} \stackrel{\mathrm{def}}{=} \begin{bmatrix} \widetilde{a}_{11} & \widetilde{a}_{12}\\\widetilde{a}_{21} & 1\end{bmatrix}, \qquad     \widetilde{\mat{B}} \stackrel{\mathrm{def}}{=} \begin{bmatrix} \widetilde{b}_{11} & \widetilde{b}_{12}\\\widetilde{b}_{21} & 1\end{bmatrix},
    \label{eq:2.3}
\end{equation}

Since the error term $k$ is defined as a common scalar from the error boxes, we can define its residual error as the product of the residual scalars from the error boxes, i.e., the terms $\widetilde{k}_a$ and $\widetilde{k}_b$ from (\ref{eq:2.2}). This can be expressed as follows:
\begin{equation}
    \widetilde{k} \stackrel{\mathrm{def}}{=} \widetilde{k}_a\widetilde{k}_b.
    \label{eq:2.4}
\end{equation}

Thus, applying a non-ideal multiline TRL calibration to a DUT can be described as follows:
\begin{equation}
    \widehat{\mat{T}}_\mathrm{dut} = \frac{1}{k}\widehat{\mat{A}}^{-1}\mat{M}_\mathrm{dut}\widehat{\mat{B}}^{-1} = \widetilde{k}\widetilde{\mat{A}}\mat{T}_\mathrm{dut}\widetilde{\mat{B}}.
    \label{eq:2.5}
\end{equation}

Ideally, the calibrated measurement of the DUT should be equal to the actual DUT. In this paper, we investigate the influence of error in the reference impedance. We assume that any residual errors are caused only by impedance variation in the line standards during multiline TRL calibration \cite{Lenk2013}. Therefore, the residual errors can be reduced to an impedance transformation error, which can be described by the following expression \cite{Marks1992}:
\begin{equation}
    \widehat{\mat{T}}_\mathrm{dut} = \underbrace{\frac{1}{1-\widetilde{\Gamma}^2}}_{\widetilde{k}}\underbrace{\begin{bmatrix} 1 & \widetilde{\Gamma}\\\widetilde{\Gamma} & 1\end{bmatrix}}_{\widetilde{\mat{A}}}\mat{T}_\mathrm{dut}\underbrace{\begin{bmatrix} 1 & -\widetilde{\Gamma}\\-\widetilde{\Gamma} & 1\end{bmatrix}}_{\widetilde{\mat{B}}},
    \label{eq:2.6}
\end{equation}
where $\widetilde{\Gamma}$ is the residual reflection coefficient, defined by:
\begin{equation}
    \widetilde{\Gamma} = \frac{Z_0 - \widetilde{Z}}{Z_0 + \widetilde{Z}},
    \label{eq:2.7}
\end{equation}

Here, $Z_0$ represents the true reference impedance, and $\widetilde{Z}$ represents the perturbed reference impedance. In an error-free scenario, $\widetilde{Z} = Z_0$, and therefore $\widetilde{\Gamma}=0$. However, when we assume the presence of random error, we can assign $\widetilde{\Gamma}$ to be a random process described by its probability distribution. In general, the exact probability distribution of $\widetilde{\Gamma}$ may be unknown. In such cases, we can assume a normal distribution, $\widetilde{\Gamma}\sim \mathcal{N}\left(0, \mat{\Sigma}_{\widetilde{\Gamma}}\right)$, where $\mat{\Sigma}_{\widetilde{\Gamma}}$ represents the covariance matrix of $\widetilde{\Gamma}$.

For an arbitrary two-port DUT, its S-parameters after calibration are determined by converting the T-parameters in \eqref{eq:2.6} to S-parameters. This results in the following:
\begin{equation}
    \widehat{\mat{S}}_\mathrm{dut} = \begin{bmatrix}
    \frac{\widetilde{\Gamma}^2S_{22} + \widetilde{\Gamma}(\mathrm{det}(\mat{S}) + 1) + S_{11}}{\widetilde{\Gamma}^2\mathrm{det}(\mat{S}) + \widetilde{\Gamma}(S_{11} + S_{22}) + 1} & \frac{S_{12}(1-\widetilde{\Gamma}^2)}{\widetilde{\Gamma}^2\mathrm{det}(\mat{S}) + \widetilde{\Gamma}(S_{11} + S_{22}) + 1}\\[5pt]
    \frac{S_{21}(1-\widetilde{\Gamma}^2)}{\widetilde{\Gamma}^2\mathrm{det}(\mat{S}) + \widetilde{\Gamma}(S_{11} + S_{22}) + 1} & \frac{\widetilde{\Gamma}^2S_{11} + \widetilde{\Gamma}(\mathrm{det}(\mat{S}) + 1) + S_{22}}{\widetilde{\Gamma}^2\mathrm{det}(\mat{S}) + \widetilde{\Gamma}(S_{11} + S_{22}) + 1}
    \end{bmatrix},
    \label{eq:2.8}
\end{equation}
where $\mathrm{det}(\mat{S}) = S_{22}S_{11} - S_{12}S_{21}$. 

To analyze the impact of $\widetilde{\Gamma}$ on the calibrated measurements, we calculate the sensitivity of $|\widehat{S}{ij}|$ with respect to the real and imaginary parts of $\widetilde{\Gamma}$. We can derive the sensitivity equations of $|\widehat{S}{ij}|$ using the Wirtinger formulation \cite{Wirtinger1927}, which relates complex-valued and real-valued differentiation as follows:
\begin{equation}
    \begin{bmatrix}
    \frac{\partial |\widehat{S}_{ij}|}{\partial \RE{\widetilde{\Gamma}}}\\[5pt]
    \frac{\partial |\widehat{S}_{ij}|}{\partial \IM{\widetilde{\Gamma}}}
    \end{bmatrix} = \begin{bmatrix}
    1 & 1\\[5pt]
    j & -j
    \end{bmatrix}\begin{bmatrix}
    \frac{\partial |\widehat{S}_{ij}|}{\partial \widetilde{\Gamma}}\\[5pt]
    \frac{\partial |\widehat{S}_{ij}|}{\partial \widetilde{\Gamma}^*}
    \end{bmatrix},
    \label{eq:2.9}
\end{equation}
where the derivatives $\frac{\partial |\widehat{S}_{ij}|}{\partial \widetilde{\Gamma}}$ and $\frac{\partial |\widehat{S}_{ij}|}{\partial \widetilde{\Gamma}^*}$ are the derivatives of $|\widehat{S}_{ij}|$ with respect to $\widetilde{\Gamma}$ and its conjugate, which are computed using the following equations:
\begin{subequations}
	\begin{align}
	    \frac{\partial |\widehat{S}_{ij}|}{\partial \widetilde{\Gamma}} &= \frac{1}{2|\widehat{S}_{ij}|}\left( \widehat{S}_{ij}^*\frac{\partial \widehat{S}_{ij}}{\partial \widetilde{\Gamma}}  + \widehat{S}_{ij}\frac{\partial \widehat{S}_{ij}^*}{\partial \widetilde{\Gamma}}\right),\label{eq:2.10a}\\[5pt]
	    \frac{\partial |\widehat{S}_{ij}|}{\partial \widetilde{\Gamma}^*} &= \frac{1}{2|\widehat{S}_{ij}|}\left( \widehat{S}_{ij}^*\frac{\partial \widehat{S}_{ij}}{\partial \widetilde{\Gamma}^*}  + \widehat{S}_{ij}\frac{\partial \widehat{S}_{ij}^*}{\partial \widetilde{\Gamma}^*}\right),\label{eq:2.10b}
	\end{align}
	\label{eq:2.10}
\end{subequations}
with $|\widehat{S}_{ij}| > 0$. Since $\widehat{S}_{ij}$ is a rational function with respect to $\widetilde{\Gamma}$, it is straightforward to show that
\begin{equation}
    \frac{\partial\widehat{S}_{ij}}{\partial \widetilde{\Gamma}^*} = \frac{\partial\widehat{S}_{ij}^*}{\partial \widetilde{\Gamma}} = 0, \qquad \frac{\partial \widehat{S}_{ij}^*}{\partial \widetilde{\Gamma}^*} = \left(\frac{\partial \widehat{S}_{ij}}{\partial \widetilde{\Gamma}}\right)^*.
    \label{eq:2.11}
\end{equation}

Accordingly, combining the results of \eqref{eq:2.10} and \eqref{eq:2.11} and inserting them into \eqref{eq:2.9}, we have the general sensitivity equations of $|\widehat{S}_{ij}|$ with respect to $\widetilde{\Gamma}$ given by
\begin{equation}
    \begin{bmatrix}
    \frac{\partial |\widehat{S}_{ij}|}{\partial \RE{\widetilde{\Gamma}}}\\[5pt]
    \frac{\partial |\widehat{S}_{ij}|}{\partial \IM{\widetilde{\Gamma}}}
    \end{bmatrix} = \begin{bmatrix}
    \RE{\frac{\widehat{S}_{ij}^*}{|\widehat{S}_{ij}|}\frac{\partial \widehat{S}_{ij}}{\partial \widetilde{\Gamma}}}
    \\[5pt]
    \IM{\frac{-\widehat{S}_{ij}^*}{|\widehat{S}_{ij}|}\frac{\partial \widehat{S}_{ij}}{\partial \widetilde{\Gamma}}}
    \end{bmatrix} = \begin{bmatrix}
    \RE{\frac{|\widehat{S}_{ij}|}{\widehat{S}_{ij}}\frac{\partial \widehat{S}_{ij}}{\partial \widetilde{\Gamma}}}
    \\[5pt]
    \IM{\frac{|\widehat{S}_{ij}|}{-\widehat{S}_{ij}}\frac{\partial \widehat{S}_{ij}}{\partial \widetilde{\Gamma}}}
    \end{bmatrix}.
    \label{eq:2.12}
\end{equation}

For calibration verification, we require a standard that is most sensitive to mismatch error. Such a standard allows us to identify the error in the reference impedance of the calibration. For instance, we can consider a one-port device described by the following expression when setting $S_{21}=S_{12}=0$ in \eqref{eq:2.8}.
\begin{equation}
    \widehat{S}_{11} = \frac{\widetilde{\Gamma} + S_{11}}{\widetilde{\Gamma}S_{11} + 1}, \qquad \frac{\partial \widehat{S}_{11}}{\partial \widetilde{\Gamma}} = \frac{1-S_{11}\widehat{S}_{11}}{\widetilde{\Gamma}S_{11} + 1}.
    \label{eq:2.13}
\end{equation}

\begin{figure}[th!]
	\centering
	\includegraphics[width=1\linewidth]{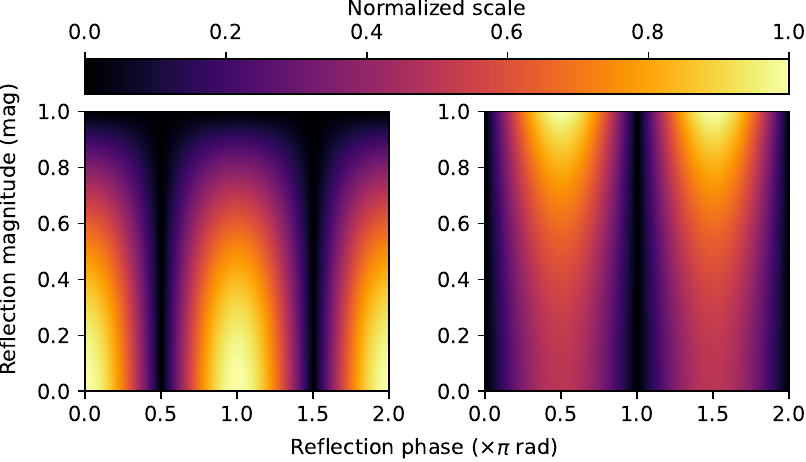}
    \caption{Plot of the sensitivity of the calibrated one-port device for $\widetilde{\Gamma}$. The left plot corresponds to $\partial |\widehat{S}{11}|/\partial \RE{\widetilde{\Gamma}}$, while the right plot corresponds to $\partial |\widehat{S}{11}|/\partial \IM{\widetilde{\Gamma}}$.}
	\label{fig:2.2}
\end{figure}

The sensitivity equations of $|\widehat{S}_{11}|$ is obtained by plugging \eqref{eq:2.13} into \eqref{eq:2.12}. Fig.~\ref{fig:2.2} depicts \eqref{eq:2.12} for an arbitrary one-port device. It is clear from Fig.~\ref{fig:2.2} that we achieve the highest sensitivity for $\RE{\widetilde{\Gamma}}$ in the regions where the phase of $S_{11}$ is at an integer multiple of $\pi$ and when it is closely matched. This evaluation also shows that we achieve the lowest sensitivity when highly reflective standards are used. The sensitivity to $\IM{\widetilde{\Gamma}}$ is greatest for $S_{11}$ measurements with standards having a phase at an integer multiple of $\pi/2$ and with high reflectivity. It is worth mentioning that for the majority of transmission lines we have $|\IM{\widetilde{\Gamma}}| \ll |\RE{\widetilde{\Gamma}}|$ (for example, see the measurements in Section~\ref{sec:5}). Therefore, a one-port device closely matched to the reference plane would be the best candidate to identify errors in the reference impedance. However, manufacturing and accurately characterizing load standards is challenging at millimeter-wave frequencies and beyond \cite{Su2021}. Therefore, using a one-port load standard is not ideal for calibration verification unless it is well characterized.

An inconvenience of using one-port devices as verification standards is the inability to assess uncertainties in transmission terms, specifically $S_{12}$ and $S_{21}$. A commonly used two-port device for calibration verification is the stepped impedance line, also known as the Beatty line \cite{EURAMET2018}. The S-parameters of an ideal stepped impedance line are given below:
\begin{equation}
    \mat{S}_\mathrm{step} = \begin{bmatrix}
    \frac{\Gamma_{nm}(e^{2\gamma l}-1)}{e^{2\gamma l} -\Gamma_{nm}^2} & \frac{e^{\gamma l}(1-\Gamma_{nm}^2)}{e^{2\gamma l} -\Gamma_{nm}^2}\\[5pt]
    \frac{e^{\gamma l}(1-\Gamma_{nm}^2)}{e^{2\gamma l} -\Gamma_{nm}^2} & \frac{\Gamma_{nm}(e^{2\gamma l}-1)}{e^{2\gamma l} -\Gamma_{nm}^2}
    \end{bmatrix},
    \label{eq:2.14}
\end{equation}
where $l$ and $\gamma$ are the length and propagation constant of the stepped line. The term $\Gamma_{nm}$ is the reflection coefficient of the impedance transition from $Z_{n}$ to $Z_{m}$, which is given by
\begin{equation}
    \Gamma_{nm} = \frac{Z_m - Z_n}{Z_m + Z_n},
    \label{eq:2.15}
\end{equation}

\begin{figure}[th!]
	\centering
	\includegraphics[width=1\linewidth]{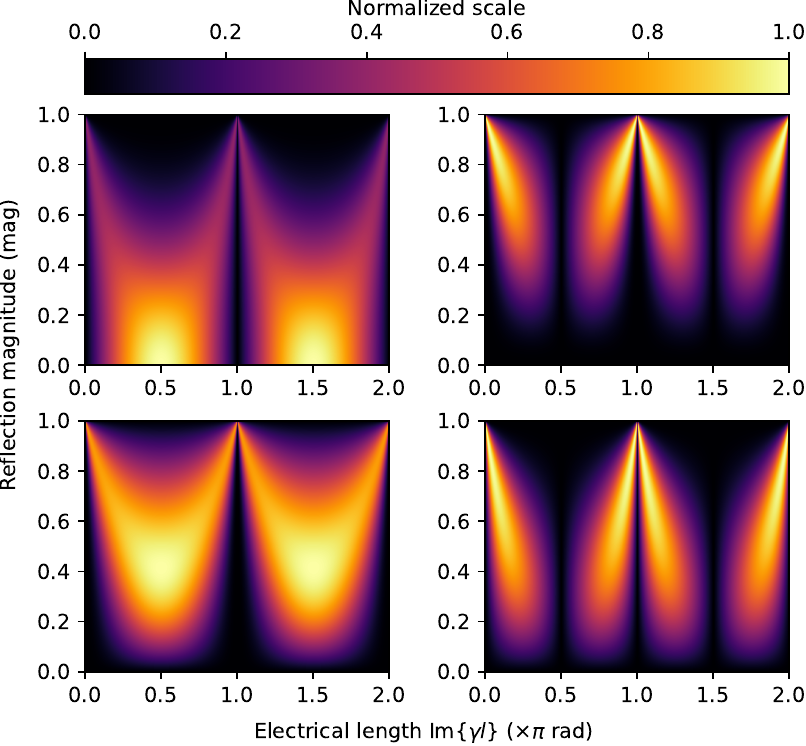}
	\caption{Plot of the sensitivity of the calibrated stepped impedance line (Beatty line) for $\widetilde{\Gamma}$. The top left and right plots correspond to $\partial |\widehat{S}_{11}|/\partial \RE{\widetilde{\Gamma}}$ and $\partial |\widehat{S}_{11}|/\partial \IM{\widetilde{\Gamma}}$, while the bottom plots correspond to $\partial |\widehat{S}_{21}|/\partial \RE{\widetilde{\Gamma}}$ and $\partial |\widehat{S}_{21}|/\partial \IM{\widetilde{\Gamma}}$, respectively.}
	\label{fig:2.3}
\end{figure}

Similar to the one-port case, we can calculate the sensitivity of the calibrated stepped impedance line by inserting \eqref{eq:2.14} into \eqref{eq:2.8}, computing the derivatives with respect to $\widetilde{\Gamma}$, and evaluating \eqref{eq:2.12}. The expressions for the derivatives are lengthy and are not presented here. However, they were assessed using the Python symbolic package \textit{SymPy} \cite{Meurer2017}. The sensitivities of $|\widehat{S}_{11}|$ and $|\widehat{S}_{21}|$ are shown in Fig.~\ref{fig:2.3}. From the figure, we see that $|\widehat{S}_{21}|$ is not sensitive to impedance variation when the stepped impedance is matched, regardless of the electrical length. This is equivalent to a line standard without any discontinuity. Due to this reason, we cannot use the line standards used in the calibration as verification standards to identify impedance errors, as they have the same impedance as the reference impedance. Generally, the sensitivity of $|\widehat{S}_{11}|$ and $|\widehat{S}_{21}|$ varies as a function of electrical length and equals zero at an integer multiple of half-wavelength. Therefore, using a single stepped impedance line cannot cover all frequencies.

The frequency limitation that we observe in the sensitivity graph shown in Fig.~\ref{fig:2.3} is similar to the frequency limitation of the TRL calibration. Therefore, we can cover a wide range of frequencies by using multiple stepped impedance standards of different lengths. To combine the results of every stepped impedance standard, we propose a second multiline TRL calibration that uses the stepped impedance lines as standards. The error boxes between the reference multiline TRL calibration and the stepped impedance calibration correspond to the impedance transformation.


%% file: Sections/Section3.tex
\section{Extraction the Reflection Coefficient of the Impedance Transition}
\label{sec:3}

\begin{figure}[th!]
	\centering
	\includegraphics[width=1\linewidth]{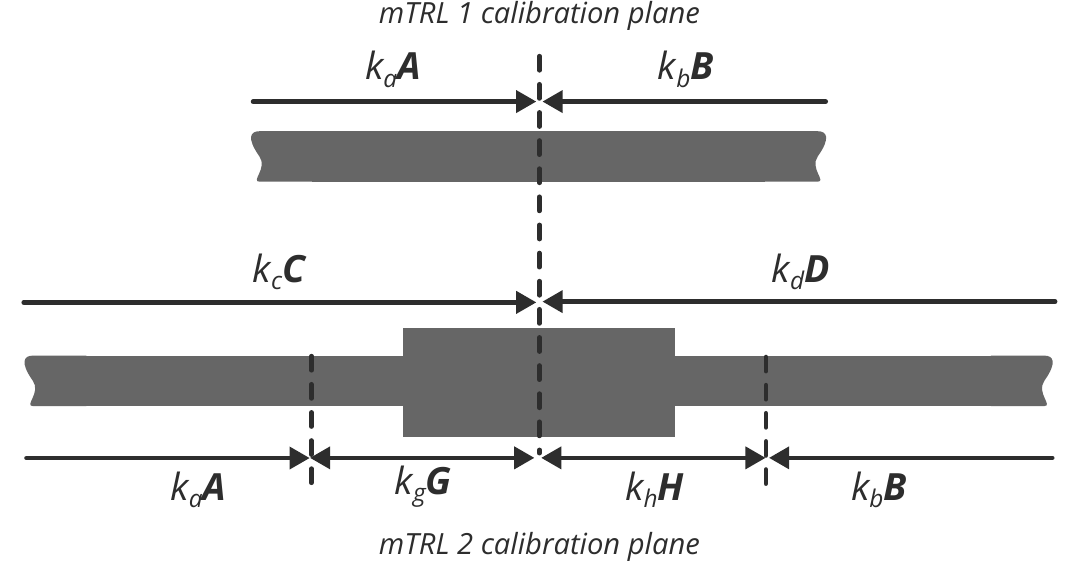}
    \caption{Illustration of a thru standard using two multiline TRL calibration kits. The letters $k_a\mat{A}$ and $k_b\mat{B}$ correspond to the error boxes of the primary TRL calibration, while the letters $k_c\mat{C}$ and $k_d\mat{D}$ correspond to the error boxes of the second TRL calibration. The letters $k_g\mat{G}$ and $k_h\mat{H}$ indicate the left and right stepped impedance segments.}
	\label{fig:3.1}
\end{figure}

The objective of our methodology is to conduct two multiline TRL calibrations using lines with different impedances. This approach enables us to evaluate the error boxes associated with the impedance transition segment between the two calibrations. From the error boxes, we can derive the reflection coefficient of the impedance transition, which serves as our validation metric later on. We begin the derivation with Fig.~\ref{fig:3.1}, which depicts an example of a microstrip thru standard with two different impedances (i.e., different trace widths).

Similar to the discussion in the previous section, we use T-parameters to describe the error box model of both multiline TRL calibrations. The error box model of the primary multiline TRL calibration is given by
\begin{equation}
	\mat{M}_1 = \underbrace{k_ak_b}_{k_1}
	\underbrace{
		\begin{bmatrix} a_{11} & a_{12}\\a_{21} & 1\end{bmatrix}}_{\mat{A}}
	\mat{T}_\mathrm{dut}
	\underbrace{\begin{bmatrix} b_{11} & b_{12}\\b_{21} & 1\end{bmatrix}}_{\mat{B}},
	\label{eq:3.1}
\end{equation}
and the error box model of the second multiline TRL calibration is given by
\begin{equation}
	\mat{M}_2 = \underbrace{k_ck_d}_{k_2}
	\underbrace{\begin{bmatrix} c_{11} & c_{12}\\c_{21} & 1\end{bmatrix}}_{\mat{C}}
	\mat{T}_\mathrm{dut}
	\underbrace{\begin{bmatrix} d_{11} & d_{12}\\d_{21} & 1\end{bmatrix}}_{\mat{D}}.
	\label{eq:3.2}
\end{equation}

Based on Fig.~\ref{fig:3.1}, we can see that the error boxes of the second multiline TRL include those of the first multiline TRL and the impedance transition segments. By performing both multiline TRL calibrations, we can determine the matrices $\mat{A}$ and $\mat{B}$ from the primary multiline TRL calibration, and the matrices $\mat{C}$ and $\mat{D}$ from the second multiline TRL calibration. It is important to note that the second multiline TRL is performed without a previous calibration, similar to the first one, but in reference to the stepped impedance. With the help of these matrices, we can calculate the T-parameters of the left stepped impedance segment as follows:
\begin{equation}
	\underbrace{\begin{bmatrix} g_{11} & g_{12}\\g_{21} & 1\end{bmatrix}}_{\mat{G}} = \frac{a_{11}-a_{21}a_{12}}{a_{11}-a_{21}c_{12}}\mat{A}^{-1}\mat{C},
	\label{eq:3.3}
\end{equation}
and the T-parameters of the right stepped impedance segment are determined as follows:
\begin{equation}
	\underbrace{\begin{bmatrix} h_{11} & h_{12}\\h_{21} & 1\end{bmatrix}}_{\mat{H}} = \frac{b_{11}-b_{12}b_{21}}{b_{11}-b_{12}d_{21}}\mat{D}\mat{B}^{-1}.
	\label{eq:3.4}
\end{equation}

Note that we have not considered the terms $k_g$ and $k_h$ in the previous equations. This is because the transition segments are reciprocal devices, i.e., $S_{12}=S_{21}$. The terms $k_g$ and $k_h$ are implicitly contained in $g_{11}$ and $h_{11}$ through the conversion relationship between S- and T-parameters. Therefore, there is no mathematical benefit in including $k_g$ and $k_h$ in the derivation.

\subsection{Modeling the Stepped Impedance Transition}

In order to accurately extract the reflection coefficient of the impedance transition, we need a model that accounts for the non-ideal parasitic effects of the structure, which are unrelated to the impedance transformation itself. For simplicity, we will focus only on the left impedance transition in the following derivation.

\begin{figure}[th!]
	\centering
	\includegraphics[width=1\linewidth]{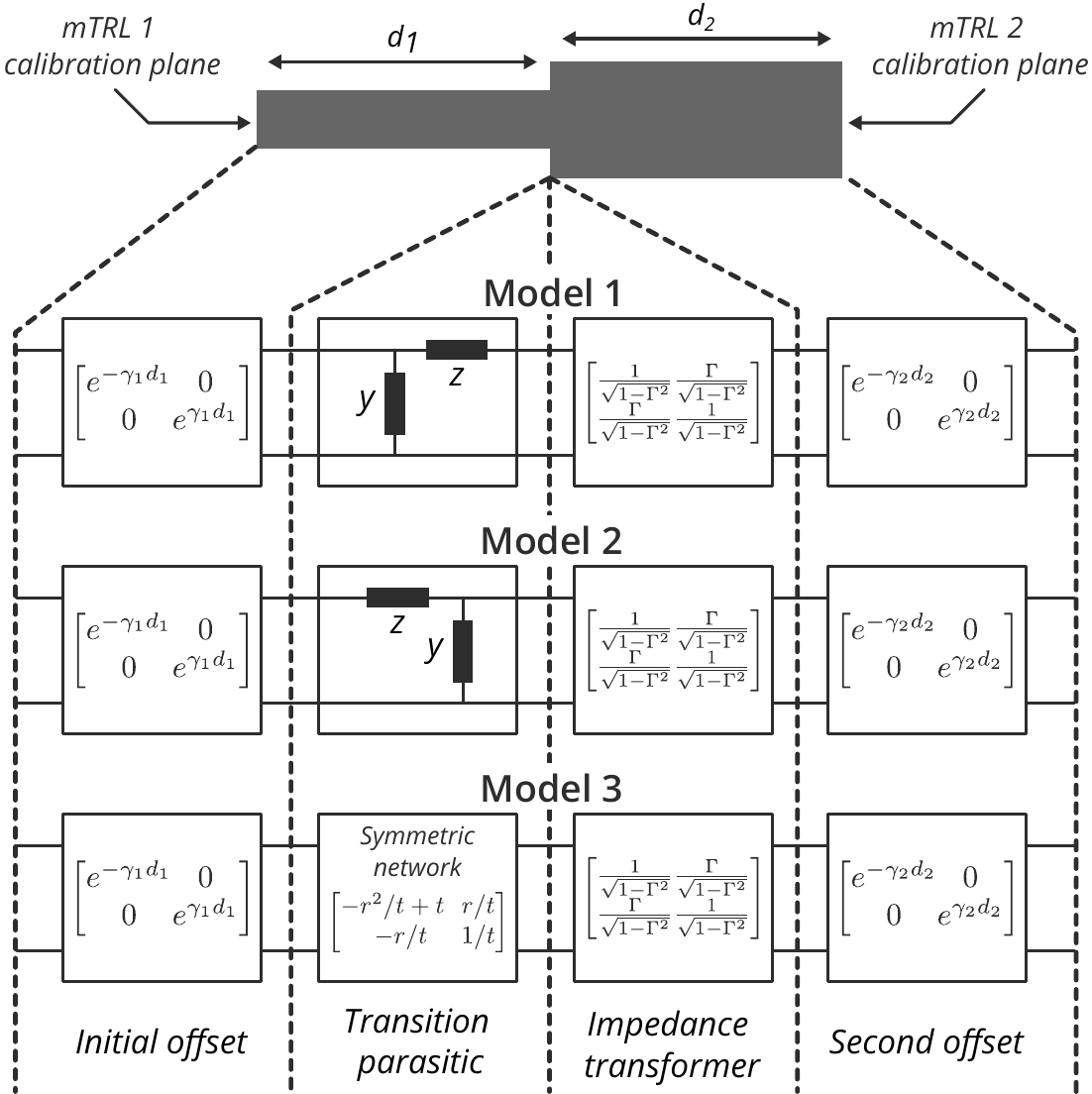}
	\caption{Proposed models for the impedance transition segment. All matrices are given as T-parameters.}
	\label{fig:3.2}
\end{figure}

To determine the reflection coefficient of the left impedance transition, we propose three possible models to characterize the parasitics of the transition, as shown in Fig.~\ref{fig:3.2}. In general, the stepped impedance segment can be divided into four blocks.
\begin{enumerate}	
	\item \textit{Initial offset:} This offset accounts for the case when the impedance discontinuity occurs at a distance from the primary multiline TRL calibration. Generally, this offset is characterized by the propagation constant $\gamma_1$ and the physical length $d_1$ of the offset.
	
	\item \textit{Transition parasitic:} The transition between impedances does not occur instantaneously. Instead, it exhibits non-ideal parasitic behavior due to the abrupt discontinuity. This can be accounted for by one of the three proposed parasitic models shown in Fig.~\ref{fig:3.2}.
	
	\item \textit{Ideal impedance transformer:} This element accounts for the actual impedance transformation, which is defined by $\Gamma_{nm}$ in \eqref{eq:2.15}. For simplicity of notation, we drop the indices, i.e., $\Gamma_{nm}=\Gamma$.
	
	\item \textit{Second offset:} The offset after the impedance transformation is an essential part of the transition design. This is because the stepped impedance cannot be realized without some length. We can determine this offset based on knowledge of the propagation constant $\gamma_2$ and the physical length $d_2$.
\end{enumerate}

It is worth mentioning that the placement of the parasitic model before or after the ideal impedance transformer is arbitrary. This is because the behavior of the parasitic transition is unknown, and scaling it with an impedance transformation makes no difference in the derivation.

To solve for the unknowns, we write the cascaded matrices of Fig.~\ref{fig:3.2} in relation to the matrix $\mat{G}$ as follows:
\begin{equation}
    k_g\mat{G} = \frac{1}{\sqrt{1-\Gamma^2}}\mat{L}_1
    \mat{P}
    \begin{bmatrix}
       1  & \Gamma\\[5pt]
       \Gamma & 1
    \end{bmatrix}\mat{L}_2,
    \label{eq:3.5}
\end{equation}
where $\mat{L}_1$ and $\mat{L}_2$ represent the left and right offsets, respectively, as shown in Fig.~\ref{fig:3.2}. The matrix $\mat{P}$ represents the parasitic transition, which is defined by one of the three models in Fig.~\ref{fig:3.2} to be equal to:
\begin{subequations}
\begin{align}
    \mat{P}^{(1)} &= \frac{1}{2}\begin{bmatrix}
        (1-y)(1-z) + 1 & (1-y)(1 + z) - 1\\[5pt]
        (1 + y)(1-z) - 1 & (1+y)(1+z) + 1
    \end{bmatrix}\label{eq:3.6a}\\[5pt]
    \mat{P}^{(2)} &= \frac{1}{2}\begin{bmatrix}
        (1-y)(1-z) + 1 & (y+1)(z-1) + 1\\[5pt]
        (y-1)(z+1) + 1 & (1+y)(1+z) + 1
    \end{bmatrix}\label{eq:3.6b}\\[5pt]
    \mat{P}^{(3)} &= \frac{1}{t}\begin{bmatrix}
        -r^2 + t^2 & r\\[5pt]
        -r & 1
    \end{bmatrix}.\label{eq:3.6c}
\end{align}
    \label{eq:3.6}
\end{subequations}

The T-parameters expressions for $\mat{P}^{(1)}$ and $\mat{P}^{(2)}$ were symbolically derived using the \textit{SymPy} package in Python \cite{Meurer2017}. This was done by converting the ABCD-parameters of the L-circuit models in Fig.~\ref{fig:3.2} into T-parameters \cite{Frickey1994}. Models 1 and 2 assume that parasitic behavior can be described using lumped elements $z$ and $y$. Model 3, given by $\mat{P}^{(3)}$, describes an arbitrary symmetric network with reflection and transmission coefficients $r$ and $t$, respectively. Mathematically, any model with one or two unknown variables can model the parasitic. However, if the model is inadequate in describing the parasitic behavior of the discontinuity, the values extracted for the reflection coefficient of the impedance transformer will incorporate errors. This is because the model cannot properly separate the effects of the parasitic from the impedance transformation. In Section~\ref{sec:3}-\ref{sec:3B}, we demonstrate this possible inadequacy with an example by incorporating line offset error, where lumped element models cause deviation in the extracted reflection coefficient of the impedance transformer.

What sets the proposed parasitic models apart from existing model approaches is their generality; they can be solved independently at each frequency. This is in contrast to the conventional lumped element approach that uses $LC$ elements \cite{Wight1974,Banerjee2004,Liu2016}, which are solved through nonlinear fitting over all frequency points. For the third model, a similar model was considered in \cite{Amakawa2019} as part of a calibration comparison method for on-wafer calibration. The method used an equivalent symmetric T-circuit with Y-parameters, which can be reformulated into an arbitrary symmetric network using T-parameters.

To solve for the unknowns ($\Gamma$ and parasitics), we need to construct three equations relating to the elements of $\mat{G}$. We can take the inverse of $\mat{L}_1$ and $\mat{L}_2$ on both sides of \eqref{eq:3.5}, as they are assumed to be known from performing both calibrations. Additionally, we can normalize the equation by the fourth element of the matrix, since $k_g$ is not needed. The resulting expression is as follows:
\begin{equation}
    \begin{bmatrix}
       \overline{g}_{11} & \overline{g}_{12}\\[5pt]
       \overline{g}_{21} & 1
    \end{bmatrix} = \begin{bmatrix}
       \frac{\Gamma p_{12}+p_{11}}{\Gamma p_{21}+p_{22}} & \frac{\Gamma p_{11}+p_{12}}{\Gamma p_{21}+p_{22}}\\[5pt]
       \frac{\Gamma p_{22}+p_{21}}{\Gamma p_{21}+p_{22}} & 1
    \end{bmatrix},
    \label{eq:3.7}
\end{equation}
where $p_{ij}$ are the elements of the parasitic matrix $\mat{P}$, and $\overline{g}_{ij}$ are defined as
\begin{subequations}
	\begin{align}
		\overline{g}_{11} &= g_{11}e^{2\gamma_1d_1+2\gamma_2d_2},\\ \overline{g}_{21} &= g_{21}e^{2\gamma_2d_2},\\
		\overline{g}_{12} &= g_{12}e^{2\gamma_1d_1},
	\end{align}
\label{eq:3.8}
\end{subequations}
with $g_{ij}$ being the elements of \eqref{eq:3.3}, the parameters $\{\gamma_1, d_1\}$ and $\{\gamma_2, d_2\}$ represent the propagation constant and offset length of the offset line, respectively. These parameters are illustrated in Fig.~\ref{fig:3.2}.

For the first and second models, the unknowns are $\Gamma$, $y$, and $z$. For the third model, the unknowns are $\Gamma$, $r$, and $t^2$. The solution for $\Gamma$ in the first and second models is given by:
\begin{equation}
	\Gamma^{(1,2)} = \pm\frac{(\overline{g}_{11}\pm\overline{g}_{21}\pm\overline{g}_{12}+1)^2-4(\overline{g}_{11}-\overline{g}_{21}\overline{g}_{12})}{(\overline{g}_{11}\pm\overline{g}_{21}\pm\overline{g}_{12}+1)^2+4(\overline{g}_{11}-\overline{g}_{21}\overline{g}_{12})},
	\label{eq:3.9}
\end{equation}
where the plus and minus signs are the solutions for the first and second models, respectively. Similarly, the reflection coefficient of the third model is given by:
\begin{equation}
	\Gamma^{(3)} = \frac{\overline{g}_{21}+\overline{g}_{12}}{\overline{g}_{11}+1}.
	\label{eq:3.10}
\end{equation}

Likewise, we can also derive a solution for the parasitic elements. For the first model, $y$ and $z$ are given as follows:
\begin{subequations}
    \begin{align}
    	y^{(1)} &= \frac{-\overline{g}_{11}+\overline{g}_{21}-\overline{g}_{12}+1}{\overline{g}_{11}+\overline{g}_{21}+\overline{g}_{12}+1}
    	\label{eq:3.11a},\\[5pt]
    	z^{(1)} &= \frac{(\overline{g}_{12}+1)^2-(\overline{g}_{11}+\overline{g}_{21})^2}{4(\overline{g}_{11}-\overline{g}_{21}\overline{g}_{12})},
    	\label{eq:3.11b}
	\end{align}
	\label{eq:3.11}
\end{subequations}
and for the second model, $y$ and $z$ are given as follows:
\begin{subequations}
    \begin{align}
    	y^{(2)} &= \frac{(\overline{g}_{12}-1)^2-(\overline{g}_{11}-\overline{g}_{21})^2}{4(\overline{g}_{11}-\overline{g}_{21}\overline{g}_{12})}
    	\label{eq:3.12a},\\[5pt]
    	z^{(2)} &= \frac{-\overline{g}_{11}-\overline{g}_{21}+\overline{g}_{12}+1}{\overline{g}_{11}-\overline{g}_{21}-\overline{g}_{12}+1}.
    	\label{eq:3.12b}
	\end{align}
	\label{eq:3.12}
\end{subequations}

Lastly, for the third model, $t^2$ and $r$ are given by:
\begin{subequations}
\begin{align}
	t^2 &= \frac{(\overline{g}_{11}-\overline{g}_{21}\overline{g}_{12})\left( (\overline{g}_{11}+1)^2-(\overline{g}_{21}+\overline{g}_{12})^2\right)}{(\overline{g}_{11}-\overline{g}_{21}\overline{g}_{12} -\overline{g}_{21}^2+1)^2}
	\label{eq:3.13a},\\[5pt]
	r &= \frac{\overline{g}_{12}-\overline{g}_{11}\overline{g}_{21}}{\overline{g}_{11}-\overline{g}_{21}\overline{g}_{12} -\overline{g}_{21}^2+1}.
	\label{eq:3.13b}
\end{align}
	\label{eq:3.13}
\end{subequations}

It should be noted that the equations for the parasitic elements of the three models are not used in the following discussion, as we are mainly concerned with $\Gamma$. However, these equations could be used as an accurate approach for the general characterization of impedance transitions of various transmission lines. Additionally, all the equations derived for the left side of the stepped impedance transition can be used for the right side by simply substituting $g_{ij}$ with the following relationships:
\begin{subequations}
	\begin{align}
		g_{11} &\longleftrightarrow h_{11},\label{eq:3.14a}\\
		g_{21} &\longleftrightarrow -h_{12},\label{eq:3.14b}\\
		g_{12} &\longleftrightarrow -h_{21}.\label{eq:3.14c}
	\end{align}
	\label{eq:3.14}
\end{subequations}

\subsection{Differences Between the Proposed Parasitic Models}
\label{sec:3B}

The models presented in Fig.~\ref{fig:3.2} are general, and their parameters are determined independently at each frequency point. In the error-free case, all three models should produce identical results for $\Gamma$, as they are exact solutions to \eqref{eq:3.7}. However, they behave differently under certain types of errors. For instance, the first and second models in Fig.~\ref{fig:3.2} are sensitive to length offset because the parasitic effects are modeled with complex impedances (i.e., $y$ and $z$). In contrast, the third model can account for any symmetrical error. In fact, the sensitivity of the first and second models to length offset can be advantageous in experimentally identifying length offset errors that could arise from probing and manufacturing tolerances.

\begin{figure}[th!]
	\centering
	\includegraphics[width=1\linewidth]{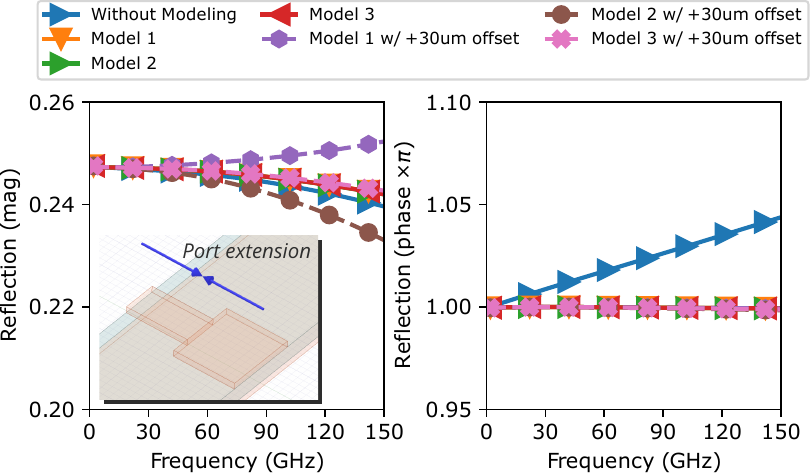}
    \caption{Extracted reflection coefficient of a simulated microstrip impedance transition segment. The port extension shifts the simulation plane to the transition. The +0.03\,mm offset error was applied at the left port.}
	\label{fig:3.3}
\end{figure}

To illustrate this point, we tested the three models to extract $\Gamma$ using electromagnetic (EM) simulation with the software ANSYS HFSS (high-frequency structure simulator). We used a microstrip line with an average impedance of approximately $53.8\,\Omega$ on one side and approximately $32.7\,\Omega$ on the other side, which results in a reflection coefficient of $-0.244$ based on \eqref{eq:2.15}. The geometric and material properties used in the simulation are the same as the mean values given in Table~\ref{tab:5.1}. A wave port extension in the simulation compensated for the offset length of the transition. Fig.~\ref{fig:3.3} shows the extracted $\Gamma$ for the three proposed models and for the case when no model is considered for the parasitic (i.e., identity matrix). We also introduced an error of +0.03\,mm offset at the $53.8\,\Omega$ line segment, whose effects can be seen in the magnitude of $\Gamma$ for the first and second models. The third model does not show a significant impact under this small offset variation. Additionally, we can see that when we do not consider a model for the parasitic, we observe deviation in both the magnitude and phase of $\Gamma$.

In general, the third model is the most reliable for capturing the reflection of the stepped impedance because it does not assume any particular type of parasitic behavior, except that it must exhibit symmetric response. For validation purposes, it is better to use all three models. If the first and second models deviate from the third model, we know that there is an error in the location of the reference plane.


%% file: Sections/Section4.tex
\section{Defining Validation Bounds}
\label{sec:4}

To validate the consistency of the reference impedance of a multiline TRL calibration, we require the extracted reflection coefficient of the impedance transition to fall within a confidence bound. For instance, the confidence bound can be set to 95\,\% coverage of a Gaussian distribution. The advantage of transmission line standards is that they can be fully characterized by knowing their cross-sectional geometry and material properties. Thus, if we know the geometric and material parameters \cite{IEC2020, BakerJarvis2005, Kato2019}, along with their uncertainties, the extracted reflection coefficient from the measurement must remain within the confidence interval derived from the propagated uncertainties.  Here, we exclude the contribution from instrumentation noise and length uncertainty, which will be discussed later in the measurements in Section~\ref{sec:5}.

To determine the confidence interval for the reflection coefficient, we propagate the uncertainties of the characteristic impedance of both transmission lines through expression \eqref{eq:2.15}. First, we need the covariance matrix of the characteristic impedance of both transmission lines. This can be calculated using the Jacobian matrix and the uncertainties of the considered parameters \cite{GuidesinMetrology2011} by means of linear propagation, which is given by:
\begin{equation}
    \mat{\Sigma}_{Z_i} = \mat{J}_{Z_i}(\bs{\mu_\theta})\mathrm{diag}\left(\begin{bmatrix} \sigma_{\theta_1}^2 & \sigma_{\theta_2}^2 & \cdots & \end{bmatrix}\right)\mat{J}_{Z_i}^T(\bs{\mu_\theta}),
    \label{eq:4.1}
\end{equation}
where $\bs{\theta}$ is the vector that contains all parameters whose mean values $\mu_{\theta_i}$ and standard uncertainties $\sigma_{\theta_i}$ are known. The Jacobian matrix $\mat{J}_{Z_i}(\bs{\theta})$ is defined as follows:
\begin{equation}
    \mat{J}_{Z_i}(\bs{\theta}) = \begin{bmatrix}
    \frac{\partial \RE{Z_i}}{\partial \theta_1} & \frac{\partial \RE{Z_i}}{\partial \theta_2} & \cdots & \\[5pt]
    \frac{\partial \IM{Z_i}}{\partial \theta_1} & \frac{\partial \IM{Z_i}}{\partial \theta_2} & \cdots &
    \end{bmatrix}.
    \label{eq:4.2}
\end{equation}

The derivatives in the Jacobian matrix can be determined from analytical models of the transmission line or directly from an EM solver. The standard uncertainty of the absolute value of the reflection coefficient is determined similarly by propagating the covariance matrices of the characteristic impedance of both calibrations from \eqref{eq:4.1} through $|\Gamma|$ by \eqref{eq:2.15}. The derived variance of $|\Gamma|$ is given as follows:
\begin{equation}
    \sigma_{|\Gamma|}^2 = \mat{J}_{|\Gamma|}(\mu_{Z_1}, \mu_{Z_2})\begin{bmatrix} \mat{\Sigma}_{Z_1} & \mat{0} \\[5pt]
    \mat{0} & \mat{\Sigma}_{Z_2}\end{bmatrix}\mat{J}^T_{|\Gamma|}(\mu_{Z_1}, \mu_{Z_2}),
    \label{eq:4.3}
\end{equation}
where $\mu_{Z_1}$ and $\mu_{Z_2}$ represent the expected values of the characteristic impedance of both transmission lines. The Jacobian matrix $\mat{J}_{|\Gamma|}(Z_1, Z_2)$ is given by:
\begin{equation}
    \mat{J}_{|\Gamma|}(Z_1, Z_2) = \begin{bmatrix}
    \frac{\partial |\Gamma|}{\partial \RE{Z_1}} & \frac{\partial |\Gamma|}{\partial \IM{Z_1}} & \frac{\partial |\Gamma|}{\partial \RE{Z_2}} & \frac{\partial |\Gamma|}{\partial \IM{Z_2}}
    \end{bmatrix}.
    \label{eq:4.4}
\end{equation}

The partial derivatives in \eqref{eq:4.4} are determined following the discussion in Section~\ref{sec:2}, using \eqref{eq:2.12}. The same derivative calculation can be applied to the phase of $\Gamma$ in a similar way.

Finally, the multiline TRL calibration is validated if the extracted reflection coefficient meets the following condition:
\begin{equation}
    \begin{array}{c}
    	\text{multiline TRL} \\ \text{Validity}
    \end{array}  = \begin{cases}
    \text{True}, & -\kappa\sigma_{|\Gamma|} \leq |\Gamma|-\mu_{|\Gamma|} \leq\kappa\sigma_{|\Gamma|}\\
    \text{False}, & \text{otherwise}
    \end{cases}
    \label{eq:4.5}
\end{equation}
where $\mu_{|\Gamma|}$ represents the expected value of the reflection coefficient, and $\kappa$ denotes the coverage factor of a Gaussian distribution. For 68\,\% coverage, $\kappa=1$. For 95\,\% coverage, $\kappa=2$, and for 99.7\,\% coverage, $\kappa=3$. If the result exceeds the confidence bounds, other types of errors may exist besides impedance variation, such as noise.


%% file: Sections/Section5.tex
\section{Experiment}
\label{sec:5}

\subsection{Measurement setup}

\begin{figure}[th!]
	\centering
	\includegraphics[width=.99\linewidth]{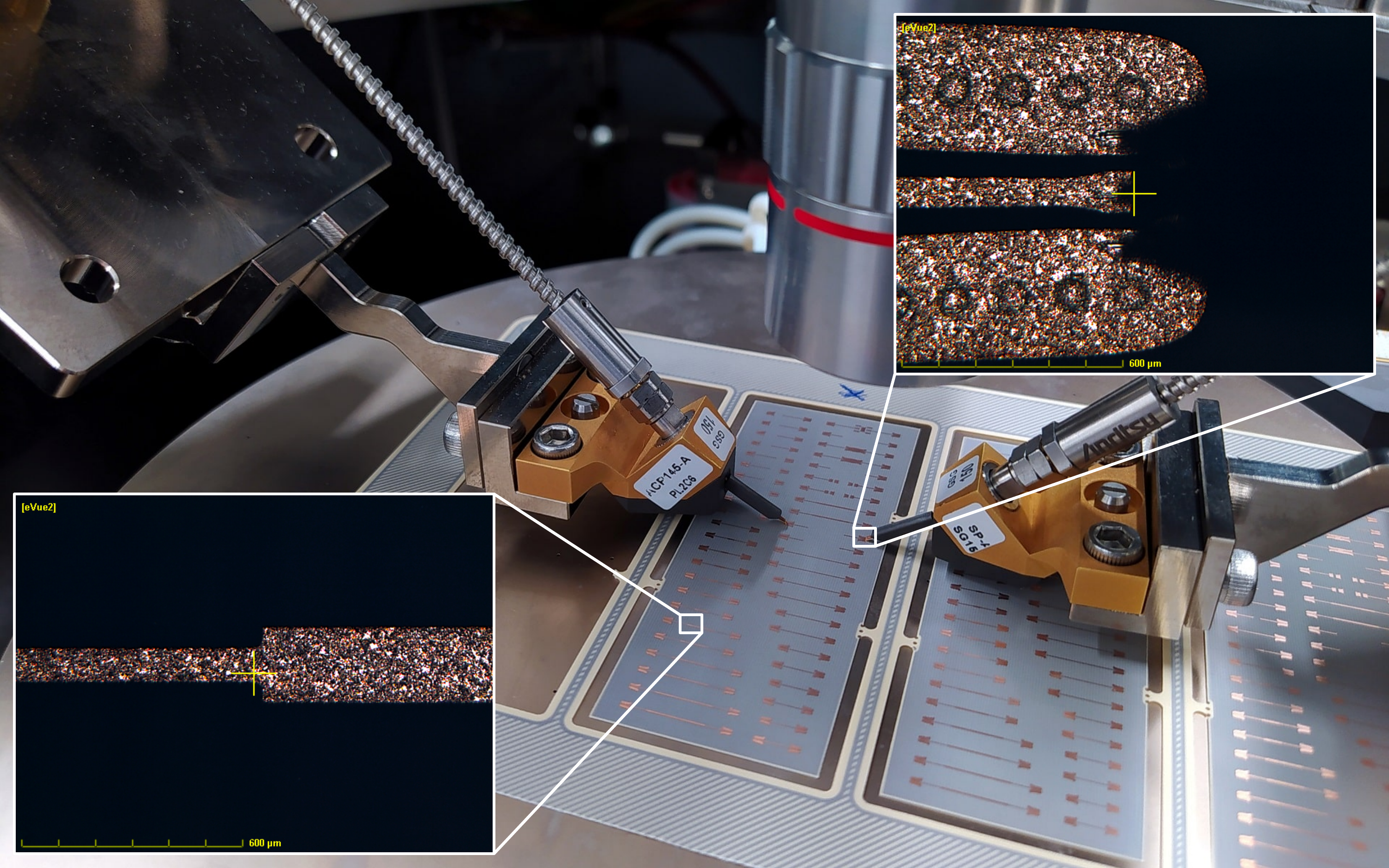}
    \caption{Photo of the measured PCB on the probe station. The bottom left inset photo shows the transition segment, whereas the top right inset photo shows the GCPW-to-microstrip pad used to transition into the microstrip structure.}
	\label{fig:5.1}
\end{figure}

The setup comprises of two multiline TRL calibration kits based on microstrip technology. As we are using a probe station with ground-signal-ground (GSG) probes for measurements, the interface pads of the microstrip lines were designed as a tapered grounded coplanar waveguide (GCPW) with optimized low-return loss based on the work in \cite{Hatab2022a}. The measurements were performed using an Anritsu VectorStar VNA with millimeter-wave extenders that support frequencies up to 150\,GHz. The probes used are ACP probes from FormFactor, with a pitch of $150\,\mu\mathrm{m}$. The probe station utilized is the semi-automatic SUMMIT200, also from FormFactor. A photo of the PCB on the probe station is shown in Fig.~\ref{fig:5.1}.

\begin{figure}[th!]
	\centering
	\subfloat[]{\includegraphics[width=.48\linewidth]{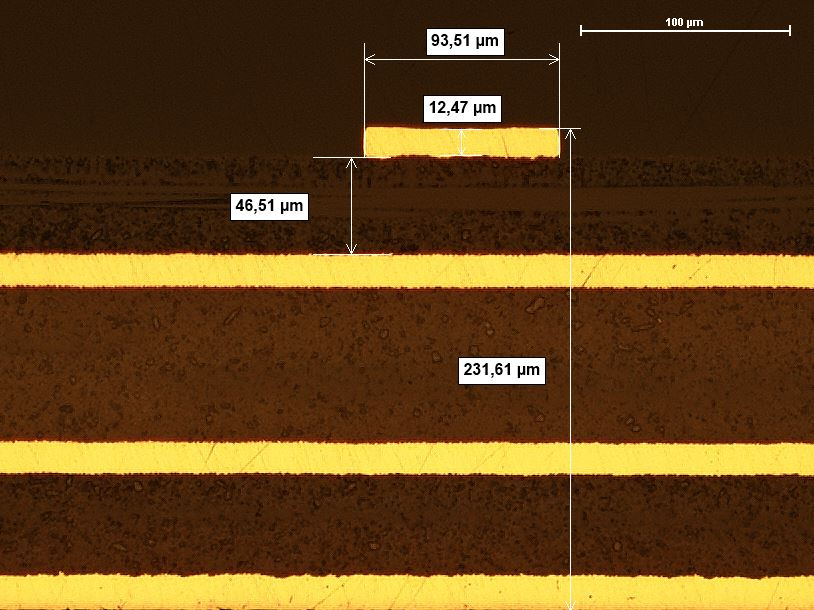}
		\label{fig:5.2a}}~
	\subfloat[]{\includegraphics[width=.48\linewidth]{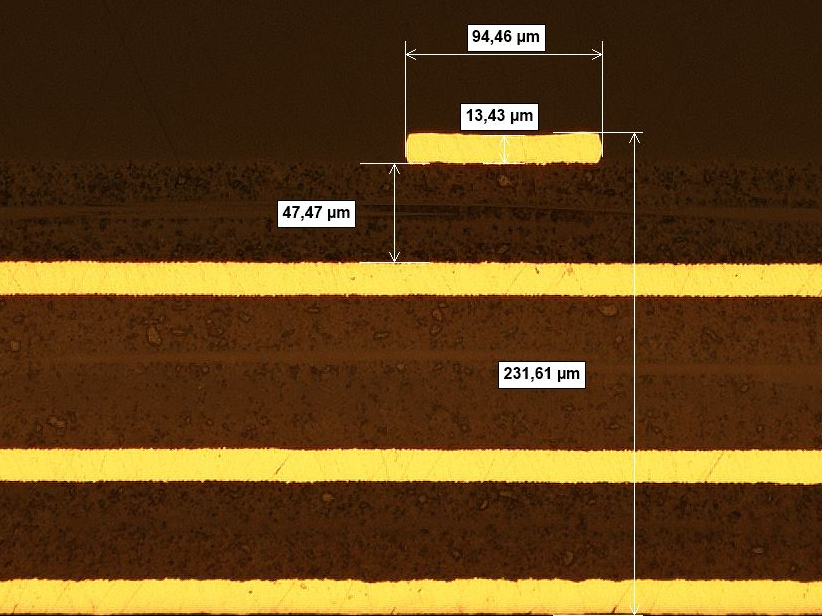}
		\label{fig:5.2b}}\\[-5pt]
	\subfloat[]{\includegraphics[width=.48\linewidth]{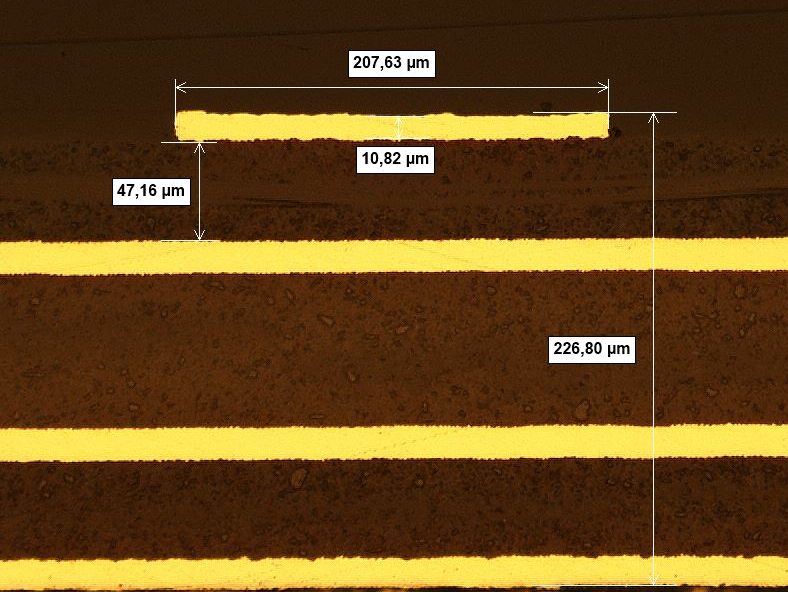}
		\label{fig:5.2c}}~
	\subfloat[]{\includegraphics[width=.48\linewidth]{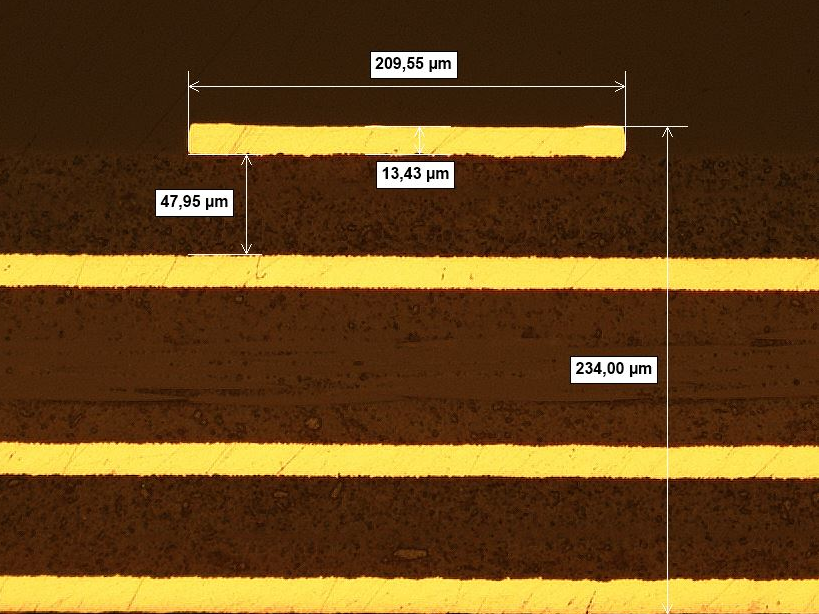}
		\label{fig:5.2d}}
	\caption{Cross-section photos of the fabricated transmission lines. (a) and (b) are the cross-section of the primary multiline TRL calibration, while (c) and (d) are from the second multiline TRL calibration (stepped lines).}
	\label{fig:5.2}
\end{figure}

The PCBs were designed using four copper layers and three dielectric substrates, with the top and bottom substrates being prepreg and the middle substrate being a core laminate. The measured structures were fabricated on the top prepreg, while the other layers were used for mechanical support. The design parameters for the microstrip lines are as follows: substrate thickness of 0.05\,mm, copper thickness (trace thickness) of 0.02\,mm, first trace width (primary multiline TRL) of $0.107\,\mathrm{mm}$, and second trace width of $0.220\,\mathrm{mm}$. From ANSYS HFSS simulation, these dimensions correspond to line standards with an average characteristic impedance across frequency of approximately $53.8\,\Omega$ and $32.7\,\Omega$, respectively. To better model the microstrip lines, a cross-sectional inspection was performed, as shown in Fig.~\ref{fig:5.2}. The fabricated dimensions are within the expected manufacturing tolerances, but differ slightly from the nominally designed values. From several cross-section photos and information from the PCB manufacturer, the estimated dimensional parameters with their uncertainties are presented in Table~\ref{tab:5.1}.
\begin{table}[ht!]
	\centering
	\caption{Microstrip line parameters used in the EM simulation to establish the expected response and the validation bounds.}
	\label{tab:5.1}
	\resizebox{0.99\columnwidth}{!}{%
		\begin{tabular}{ccccccc}
			\toprule
			\begin{tabular}[c]{@{}c@{}}Trace \\ width 1\\ (mm)\end{tabular} &
			\begin{tabular}[c]{@{}l@{}}Trace\\ width 2\\ (mm)\end{tabular} &
			\begin{tabular}[c]{@{}c@{}}Trace\\ thickness\\ (mm)\end{tabular} &
			\multicolumn{1}{c}{\begin{tabular}[c]{@{}c@{}}Substrate\\ thickness\\ (mm)\end{tabular}} &
			\begin{tabular}[c]{@{}c@{}}Dielectric \\ constant\\ (1)\end{tabular} &
			\begin{tabular}[c]{@{}c@{}}Loss\\ tangent\\ (1)\end{tabular} &
			\begin{tabular}[c]{@{}c@{}}Copper\\ conductivity\\ (Ms/m)\end{tabular} \\ \midrule
			\begin{tabular}[c]{@{}c@{}}$0.094$\\ $\pm 0.015$\end{tabular} &
			\begin{tabular}[c]{@{}l@{}}$0.209$\\ $\pm 0.015$\end{tabular} &
			\begin{tabular}[c]{@{}c@{}}$0.013$\\ $\pm 0.007$\end{tabular} &
			\begin{tabular}[c]{@{}l@{}}$0.048$\\ $\pm 0.002$\end{tabular} &
			\begin{tabular}[c]{@{}c@{}}$3.04$\\ $\pm 0.42$\end{tabular} &
			\begin{tabular}[c]{@{}c@{}}$0.0022$\\ $\pm 10\%$\end{tabular} &
			\begin{tabular}[c]{@{}c@{}}$58$\\ $\pm 10\%$\end{tabular} \\ \bottomrule
		\end{tabular}%
	}
\end{table}

For the copper foil, we assumed ideal conductivity since the traces are without surface finish, to which we assumed a 10\% uncertainty for demonstration purposes. For the dielectric substrate, we used the Megtron 7 R-5680(N) prepreg substrate from Panasonic with a fiberglass cloth style 1027 and a resin content of 77\%. The datasheet of Megtron 7 \cite{Megtron} provides typical values of the dielectric constant and loss tangent that were derived using the balanced type circular disk resonance method \cite{IEC2020}. The values for dielectric constant and loss tangent in Table~\ref{tab:5.1} were obtained from the datasheet \cite{Megtron} by averaging the values across frequency. In the EM simulation using Ansys HFSS, these values were treated as frequency-independent because we needed to calculate derivatives to propagate uncertainty, as Ansys HFSS only allows for specifying a single variable to each considered quantity when computing derivatives. Since there were no available uncertainties for the measured quantities, we assumed a 10\% uncertainty in the loss tangent for the sake of demonstration. However, we considered the uncertainty of the dielectric constant by accounting for random placement of the microstrip line above the substrate. In some cases, the microstrip might be located on a fiberglass weave, while in other cases, it might be located on the epoxy resin. Generally, fiberglass has a higher dielectric constant than epoxy resin. From the reference \cite{Scheer2020}, the ``low Dk glass'' Panasonic uses is expected to have a dielectric constant around 5. As a result, we can infer the dielectric constant of the epoxy resin by using the resin filling ratio of 77\% \cite{Megtron}, which leads to the epoxy resin having a dielectric constant of around 2.5. Therefore, we can estimate the standard uncertainty in the dielectric constant seen by the microstrip lines using these maximum and minimum values as the 99.7\% coverage of a Gaussian distribution. Under these conditions, we get a standard deviation for the dielectric constant by the below expression \cite{Shiffler1980}.
\begin{equation}
	\sigma_{\epsilon_r} \approx \frac{1}{6}\left( \mathrm{max}(\epsilon_r) - \mathrm{min}(\epsilon_r)\right) = \frac{5-2.5}{6} \approx 0.42.
	\label{eq:5.1}
\end{equation}

Both multiline TRL calibrations use six microstrip lines with lengths (relative to the first line) of $\{0,\, 0.5,\, 1,\, 3,\, 5,\, 6.5\}\,\mathrm{mm}$. The reference planes were set to the middle of the thru structure for both calibration kits (similar to the illustration in Fig.~\ref{fig:3.1}). The offset length of the impedance transition was chosen to be 0.5\,mm on both sides (that is, $d_1=d_2=0.5\,\mathrm{mm}$ in Fig.~\ref{fig:3.2}). The reflect standard was implemented as offset-open by setting the probes floating, with an offset of $-5.3\,\mathrm{mm}$.

\subsection{Results and discussion}

The measurements were taken as direct wave parameters, and corresponding S-parameters were computed. To account for noise in our analysis, we recorded 25 frequency sweeps of each measured standard to form an estimate for the covariance of the measurements due to noise from the VNA. The measurement was conducted in the frequency range from 1\,GHz to 150\,GHz with a power level of -10\,dBm and an IF (intermediate frequency) bandwidth of 100\,Hz. The data were processed offline in Python using the package \textit{scikit-rf} \cite{Arsenovic2022}. The multiline TRL algorithm used is based on \cite{Hatab2022}, and the uncertainty propagation through the calibration is based on the approach of \cite{Hatab2022b,Hatab2023}. 

The validation bounds were determined by EM simulation using the software ANSYS HFSS, which also provides calculation of derivatives while performing the simulation \cite{Vardapetyan2008}. The values listed in Table~\ref{tab:5.1} were used for the simulation and defining the validation bounds.                                                                                                                                                                                                                                                                                                                                                                                                                      In the uncertainty budget, we took into account the noise with help of the estimated sample covariance matrix and the uncertainty in the length of the line standards. We approximated the latter to be $50\,\mu\mathrm{m}$, which includes both the length uncertainty of the standards due to manufacturing and the repeatability of the probe contact location.

\begin{figure}[th!]
	\centering
	\subfloat[]{\includegraphics[width=1\columnwidth]{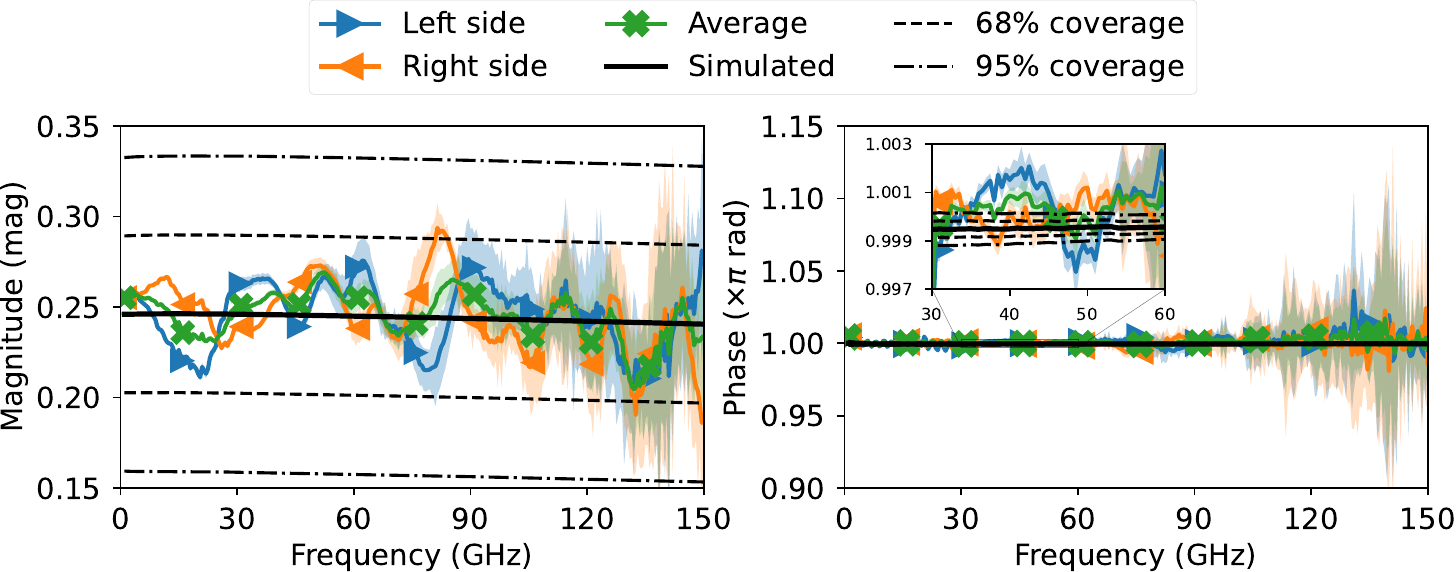}\label{fig:5.3a}}\\\vspace{-6pt}
	\subfloat[]{\includegraphics[width=1\columnwidth]{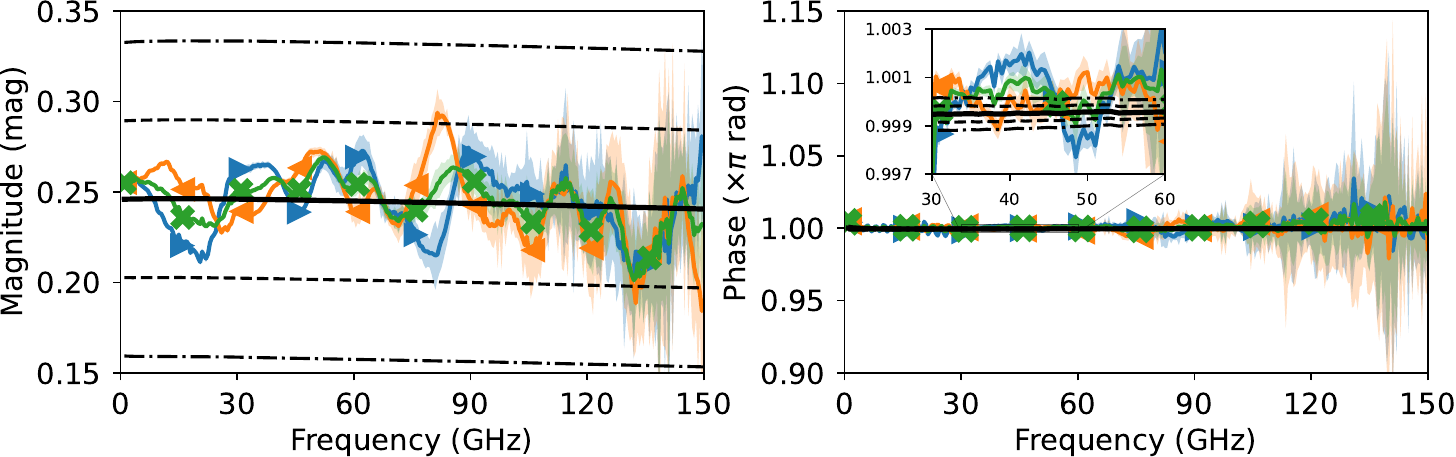}\label{fig:5.3b}}\\\vspace{-6pt}
	\subfloat[]{\includegraphics[width=1\columnwidth]{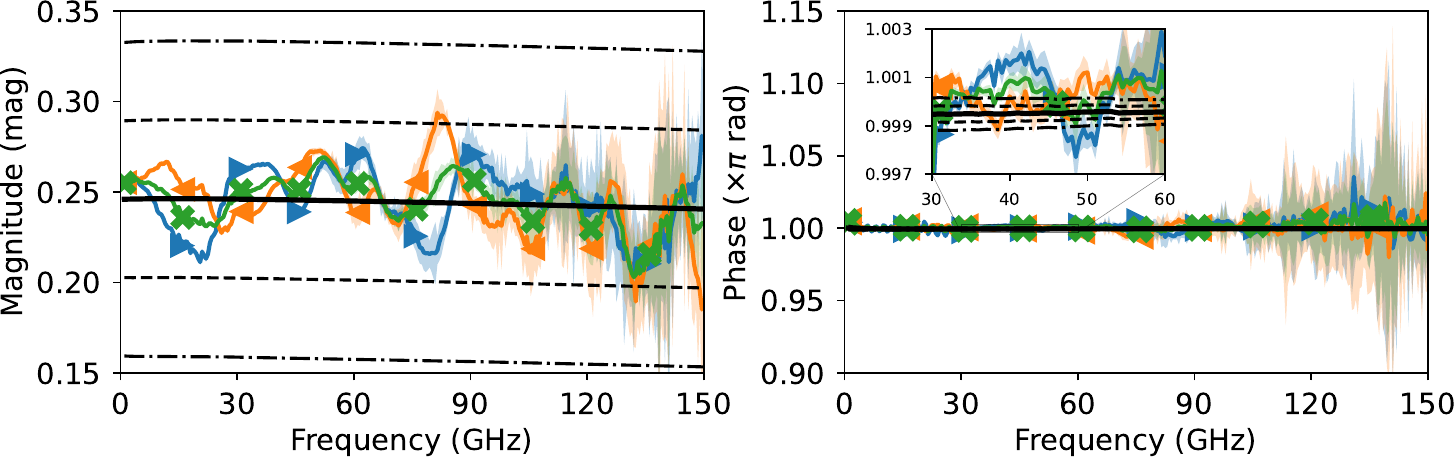}\label{fig:5.3c}}
	\caption{The measured reflection coefficient of the impedance transition based on all three models from Fig.~\ref{fig:3.2}: (a) Model 1, (b) Model 2, and (c) Model 3. The validation bounds only include cross-sectional variation that corresponds to the uncertainty values presented in Table~\ref{tab:5.1}. The uncertainty bounds around the measurements are due to propagated noise and length uncertainty through the calibration, and are reported as $95\%$ coverage.}
	\label{fig:5.3}
\end{figure}

In Fig.~\ref{fig:5.3}, we present the results of the extracted reflection of the impedance transition from both sides. The figure includes all three models, as well as the average value obtained by combining the results from both sides. The results indicate that all models produce identical results in the mean value, revealing no deterministic length offset error. However, one can observe small difference in the uncertainty bounds between the models in the magnitude response, which is related to the sensitivity of the models to length offset. 

In general, the magnitude response of the extracted reflection coefficient remains within the 95\% interval and mostly within the 68\% interval. On the other hand, the phase component has an average value of $\pi$, but is mostly impacted by noise. In general, the phase of the reflection (or equivalent the imaginary part) is mostly impacted by noise, with the confidence interval due to cross-sectional variation being much smaller than the influence of noise. For the measured magnitude, noise and length uncertainty have a more apparent impact at higher frequencies ($>100\,\mathrm{GHz}$). Therefore, impedance mismatch due to cross-sectional variation has an influence across all frequencies, but starting at higher frequencies, noise becomes prominent as well.

While the reflection measurement of the impedance transition and the validation bounds are useful indications of the calibration quality, we can also quantify the amount of error in terms of Ohms by considering the error in the characteristic impedance of the stepped line. First, we rewrite the reflection of the impedance transition in terms of normalized impedance as follows:
\begin{equation}
	\Gamma = \frac{Z_\mathrm{step}-Z_\mathrm{ref}}{Z_\mathrm{step}+Z_\mathrm{ref}} \quad\Longrightarrow\quad Z^\prime = \frac{Z_\mathrm{step}}{Z_\mathrm{ref}} = \frac{1+\Gamma}{1-\Gamma}
	\label{eq:5.2}
\end{equation}

Then, we define the error in the measured impedance of the stepped line as follows:
\begin{equation}
	\Delta Z = (Z^\prime_\mathrm{meas} - Z^\prime_\mathrm{ideal})Z_\mathrm{ref}
	\label{eq:5.3}
\end{equation}
where $Z^\prime_\mathrm{meas}$ represents the normalized impedance computed from the measurement of $\Gamma$, while $Z^\prime_\mathrm{ideal}$ is the expected normalized impedance, which can be determined from simulation. $Z_\mathrm{ref}$ is the expected characteristic impedance of the matched multiline TRL calibration, and can also be estimated through simulation.

\begin{figure}[th!]
	\centering
	\includegraphics[width=1\linewidth]{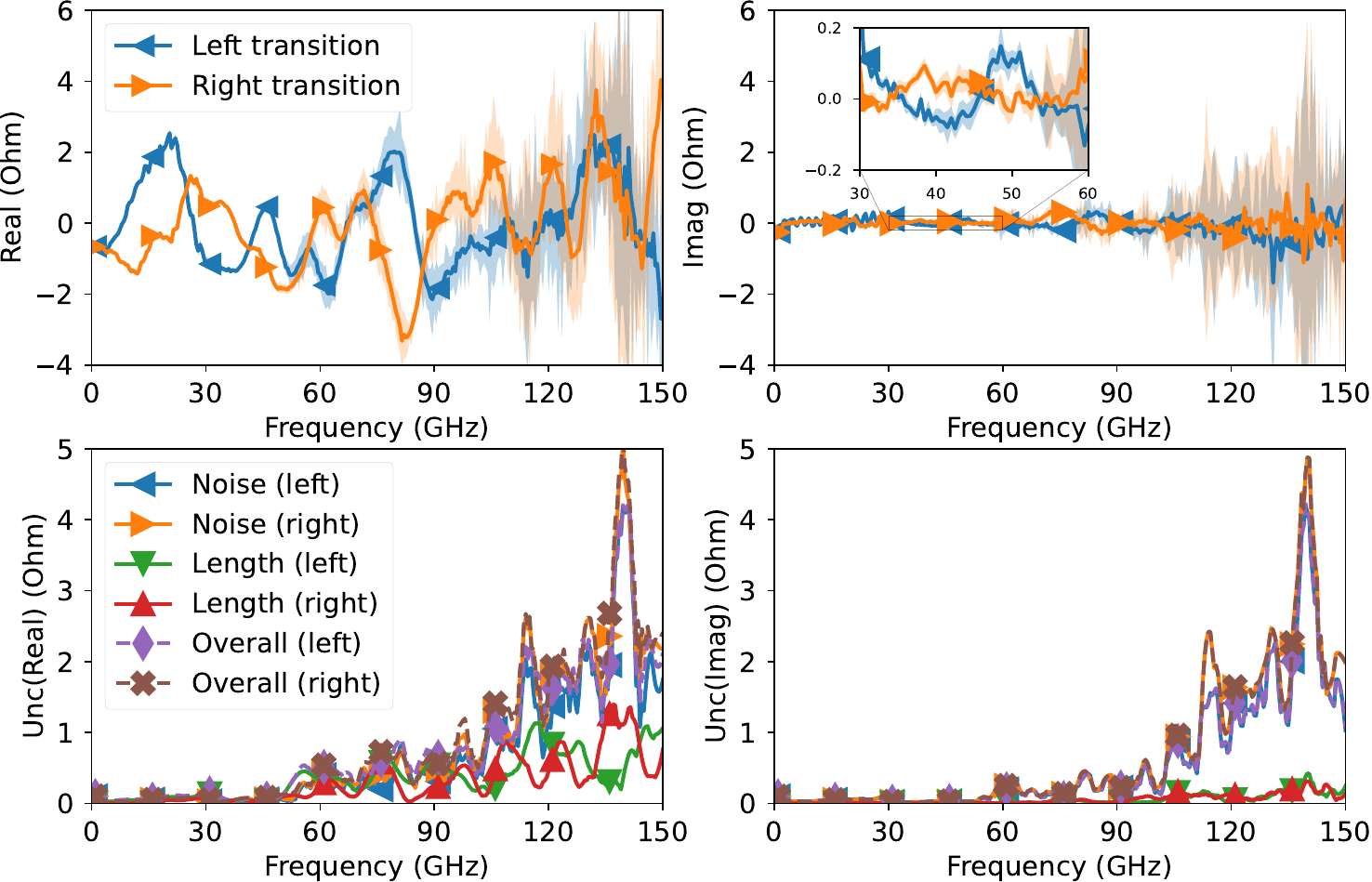}
    \caption{Error in the characteristic impedance of the stepped line based on ``Model 3'', and its associated uncertainty budget due to noise and length uncertainty. Other models are not shown as they produce similar results with small increase in uncertainty contribution due to length. The uncertainties are reported as $95\%$ coverage. The curves in the uncertainty budget graph are filtered for better readability using a Savitzky-Golay filter \cite{Savitzky1964} with a window size of 9 and a polynomial order of 2.}
	\label{fig:5.4}
\end{figure}

In Fig.~\ref{fig:5.4}, we present the error in impedance by using the simulated reflection coefficient of the impedance transition to compute the ideal normalized impedance and the simulated characteristic impedance of the reference line. The figure shows that the real part of the error exhibits impedance variation of $\pm4\,\Omega$, which falls within typical values of controlled impedance design in the PCB industry. However, after $100\,\mathrm{GHz}$, noise from the VNA dominates the variation. This is also reflected in the uncertainty budget, where noise becomes the dominant cause of variation in the real part at higher frequencies. For the imaginary part, noise is the primary cause of variation.


%% file: Sections/Section6.tex
\section{Conclusion}
\label{sec:6}

We have proposed a method to validate the reference impedance accuracy of multiline TRL calibration. To achieve this, we performed an additional multiline TRL calibration using multiple stepped impedance standards of different lengths. We then extracted the reflection coefficient of the impedance transition structure between the two multiline TRL calibrations, while accounting for the parasitic effects of the abrupt impedance change, and used it as a broadband validation metric. Our method can assess the accuracy of the reference impedance across a wide range of frequencies without explicitly measuring the characteristic impedance of the line standards.

The reflection coefficient of the impedance transition generally has a flat frequency response, even for quasi-TEM transmission lines such as microstrip lines, as demonstrated in this paper. As a result, it is an ideal validation metric that is easy to interpret. One disadvantage of our method is that it requires the measurements of another complete set of multiline TRL standards, which can be a laborious task if done manually. The upside is that we do not need to pre-characterize any standards. One way the proposed method could be improved is by using it for on-wafer applications with fully motorized probes. This allows for an automated measurement process, eliminating the need for user intervention and reducing errors that might otherwise be introduced by the user, such as those related to probe contact repeatability.


%% file: main.bbl
\begin{thebibliography}{10}
\providecommand{\url}[1]{#1}
\csname url@samestyle\endcsname
\providecommand{\newblock}{\relax}
\providecommand{\bibinfo}[2]{#2}
\providecommand{\BIBentrySTDinterwordspacing}{\spaceskip=0pt\relax}
\providecommand{\BIBentryALTinterwordstretchfactor}{4}
\providecommand{\BIBentryALTinterwordspacing}{\spaceskip=\fontdimen2\font plus
\BIBentryALTinterwordstretchfactor\fontdimen3\font minus
  \fontdimen4\font\relax}
\providecommand{\BIBforeignlanguage}[2]{{%
\expandafter\ifx\csname l@#1\endcsname\relax
\typeout{** WARNING: IEEEtran.bst: No hyphenation pattern has been}%
\typeout{** loaded for the language `#1'. Using the pattern for}%
\typeout{** the default language instead.}%
\else
\language=\csname l@#1\endcsname
\fi
#2}}
\providecommand{\BIBdecl}{\relax}
\BIBdecl
\renewcommand{\BIBentryALTinterwordstretchfactor}{4}

\bibitem{Marks1991}
R.~Marks, ``A multiline method of network analyzer calibration,'' \emph{IEEE
  Transactions on Microwave Theory and Techniques}, vol.~39, no.~7, pp.
  1205--1215,  1991, doi:
  \href{http://dx.doi.org/10.1109/22.85388}{10.1109/22.85388}.

\bibitem{Ye2017}
L.~Ye, C.~Li, X.~Sun, S.~Jin, B.~Chen, X.~Ye, and J.~Fan, ``Thru-reflect-line
  calibration technique: Error analysis for characteristic impedance variations
  in the line standards,'' \emph{IEEE Transactions on Electromagnetic
  Compatibility}, vol.~59, no.~3, pp. 779--788,  2017, doi:
  \href{http://dx.doi.org/10.1109/TEMC.2016.2623813}{10.1109/TEMC.2016.2623813}.

\bibitem{Hyunji2021}
H.~Koo, M.~Salter, N.-W. Kang, N.~Ridler, and Y.-P. Hong, ``Uncertainty of
  s-parameter measurements on pcbs due to imperfections in the trl line
  standard,'' \emph{Journal of Electromagnetic Engineering and Science},
  vol.~21, no.~5, pp. 369--378,  2021, doi:
  \href{http://dx.doi.org/10.26866/jees.2021.5.r.45}{10.26866/jees.2021.5.r.45}.

\bibitem{Lenk2013}
F.~Lenk, R.~Doerner, and A.~Rumiantsev, ``Sensitivity analysis of s-parameter
  measurements due to calibration standards uncertainty,'' \emph{IEEE
  Transactions on Microwave Theory and Techniques}, vol.~61, no.~10, pp.
  3800--3807,  2013, doi:
  \href{http://dx.doi.org/10.1109/TMTT.2013.2279774}{10.1109/TMTT.2013.2279774}.

\bibitem{Zeier2018}
M.~Zeier, J.~Hoffmann, P.~Hürlimann, J.~Rüfenacht, D.~Stalder, and
  M.~Wollensack, ``Establishing traceability for the measurement of scattering
  parameters in coaxial line systems,'' \emph{Metrologia}, vol.~55, no.~1, pp.
  S23--S36,  jan 2018, doi:
  \href{http://dx.doi.org/10.1088/1681-7575/aaa21c}{10.1088/1681-7575/aaa21c}.

\bibitem{Ridler2019}
N.~M. Ridler, R.~G. Clarke, C.~Li, and M.~J. Salter, ``Strategies for traceable
  submillimeter-wave vector network analyzer,'' \emph{IEEE Transactions on
  Terahertz Science and Technology}, vol.~9, no.~4, pp. 392--398,  2019, doi:
  \href{http://dx.doi.org/10.1109/TTHZ.2019.2911870}{10.1109/TTHZ.2019.2911870}.

\bibitem{Ridler2021}
N.~M. Ridler, S.~Johny, M.~J. Salter, X.~Shang, W.~Sun, and A.~Wilson,
  ``Establishing waveguide lines as primary standards for scattering parameter
  measurements at submillimetre wavelengths,'' \emph{Metrologia}, vol.~58,
  no.~1, p. 015015,  jan 2021, doi:
  \href{http://dx.doi.org/10.1088/1681-7575/abd371}{10.1088/1681-7575/abd371}.

\bibitem{Arz2019}
U.~Arz, K.~Kuhlmann, T.~Dziomba, G.~Hechtfischer, G.~N. Phung, F.~J.
  Schmückle, and W.~Heinrich, ``Traceable coplanar waveguide calibrations on
  fused silica substrates up to 110 ghz,'' \emph{IEEE Transactions on Microwave
  Theory and Techniques}, vol.~67, no.~6, pp. 2423--2432,  2019, doi:
  \href{http://dx.doi.org/10.1109/TMTT.2019.2908857}{10.1109/TMTT.2019.2908857}.

\bibitem{Arz2019a}
U.~Arz \emph{et~al.}, ``\BIBforeignlanguage{en}{Best practice guide for planar
  s-parameter measurements using vector network analysers : Empir - 14ind02
  planarcal},''  2019, doi:
  \href{http://dx.doi.org/10.7795/530.20190424B}{10.7795/530.20190424B}.

\bibitem{Shlepnev2014}
Y.~Shlepnev and C.~Nwachukwu, ``Modelling jitter induced by fibre weave effect
  in pcb dielectrics,'' in \emph{2014 IEEE International Symposium on
  Electromagnetic Compatibility (EMC)},  2014, doi:
  \href{http://dx.doi.org/10.1109/ISEMC.2014.6899078}{10.1109/ISEMC.2014.6899078}.
  pp. 803--808.

\bibitem{Chen2019}
B.~Chen, R.~Yao, H.~Wang, K.~Geng, and J.~Li, ``Effect of fiber weave structure
  in printed circuit boards on signal transmission characteristics,''
  \emph{Applied Sciences}, vol.~9, no.~2,  2019, doi:
  \href{http://dx.doi.org/10.3390/app9020353}{10.3390/app9020353}.

\bibitem{Lau2019}
I.~Lau \emph{et~al.}, ``Influence of the pcb manufacturing process on the
  measurement error of planar relative permittivity sensors up to 100 ghz,''
  \emph{IEEE Transactions on Microwave Theory and Techniques}, vol.~67, no.~7,
  pp. 2793--2804,  2019, doi:
  \href{http://dx.doi.org/10.1109/TMTT.2019.2910114}{10.1109/TMTT.2019.2910114}.

\bibitem{Sepaintner2020}
F.~Sepaintner, A.~Scharl, F.~Röhrl, W.~Bogner, and S.~Zorn, ``Characterization
  and production of pcb structures with increased ratio of electromagnetic
  field in air,'' \emph{IEEE Transactions on Microwave Theory and Techniques},
  vol.~68, no.~6, pp. 2134--2143,  2020, doi:
  \href{http://dx.doi.org/10.1109/TMTT.2020.2983934}{10.1109/TMTT.2020.2983934}.

\bibitem{Manukovsky2019}
A.~Manukovsky and Y.~Shlepnev, ``Measurement-assisted extraction of pcb
  interconnect model parameters with fabrication variations,'' in \emph{2019
  IEEE 28th Conference on Electrical Performance of Electronic Packaging and
  Systems (EPEPS)}.\hskip 1em plus 0.5em minus 0.4em\relax  {IEEE}, oct 2019,
  doi:
  \href{http://dx.doi.org/10.1109/EPEPS47316.2019.193228}{10.1109/EPEPS47316.2019.193228}.
  pp. 1--3.

\bibitem{Williams1991}
D.~F. Williams, R.~B. Marks, and A.~Davidson, ``Comparison of on-wafer
  calibrations,'' in \emph{38th ARFTG Conference Digest}, vol.~20,  1991, doi:
  \href{http://dx.doi.org/10.1109/ARFTG.1991.324040}{10.1109/ARFTG.1991.324040}.
  pp. 68--81.

\bibitem{Rumiantsev2006}
A.~Rumiantsev, R.~Doerner, and S.~Thies, ``Calibration standards verification
  procedure using the calibration comparison technique,'' in \emph{2006
  European Microwave Conference},  2006, doi:
  \href{http://dx.doi.org/10.1109/EUMC.2006.281416}{10.1109/EUMC.2006.281416}.
  pp. 489--491.

\bibitem{Williams2001}
D.~Williams, U.~Arz, and H.~Grabinski, ``Characteristic-impedance measurement
  error on lossy substrates,'' \emph{IEEE Microwave and Wireless Components
  Letters}, vol.~11, pp. 299--301,  2001, doi:
  \href{http://dx.doi.org/10.1109/7260.933777}{10.1109/7260.933777}.

\bibitem{Galatro2017}
L.~Galatro and M.~Spirito, ``Millimeter-wave on-wafer trl calibration employing
  3-d em simulation-based characteristic impedance extraction,'' \emph{IEEE
  Transactions on Microwave Theory and Techniques}, vol.~65, no.~4, pp.
  1315--1323,  2017, doi:
  \href{http://dx.doi.org/10.1109/TMTT.2016.2609413}{10.1109/TMTT.2016.2609413}.

\bibitem{EURAMET2018}
\BIBentryALTinterwordspacing
\emph{Guidelines on the Evaluation of Vector Network Analysers (Calibration
  Guide No. 12 Version 3.0)}, EURAMET, Mar. 2018. [Online]. Available:
  \url{https://www.euramet.org/publications-media-centre/calibration-guidelines}
\BIBentrySTDinterwordspacing

\bibitem{Marks1992}
R.~Marks and D.~Williams, ``\BIBforeignlanguage{en}{A general waveguide circuit
  theory},'' \emph{\BIBforeignlanguage{en}{Journal of Research (NIST JRES),
  National Institute of Standards and Technology, Gaithersburg, MD}}, no.~97,
  1992, doi:
  \href{http://dx.doi.org/10.6028/jres.097.024}{10.6028/jres.097.024}.

\bibitem{Wirtinger1927}
W.~Wirtinger, ``Zur formalen theorie der funktionen von mehr komplexen
  veränderlichen,'' \emph{Mathematische Annalen}, vol.~97, no.~1, pp.
  357--375,  1927, doi:
  \href{http://dx.doi.org/10.1007/BF01447872}{10.1007/BF01447872}.

\bibitem{Su2021}
J.~Su, J.~Wang, F.~Wang, and L.~Sun, ``Realization of accurate load impedance
  characterization for on-wafer trm calibration,'' \emph{Frontiers in Physics},
  vol.~8,  2021, doi:
  \href{http://dx.doi.org/10.3389/fphy.2020.595732}{10.3389/fphy.2020.595732}.

\bibitem{Meurer2017}
A.~Meurer \emph{et~al.}, ``Sympy: symbolic computing in python,'' \emph{PeerJ
  Computer Science}, vol.~3, p. e103,  Jan. 2017, doi:
  \href{http://dx.doi.org/10.7717/peerj-cs.103}{10.7717/peerj-cs.103}.

\bibitem{Frickey1994}
D.~Frickey, ``Conversions between s, z, y, h, abcd, and t parameters which are
  valid for complex source and load impedances,'' \emph{IEEE Transactions on
  Microwave Theory and Techniques}, vol.~42, no.~2, pp. 205--211,  1994, doi:
  \href{http://dx.doi.org/10.1109/22.275248}{10.1109/22.275248}.

\bibitem{Wight1974}
J.~Wight, O.~Jain, W.~Chudobiak, and V.~Makios, ``Equivalent circuits of
  microstrip impedance discontinuities and launchers,'' \emph{IEEE Transactions
  on Microwave Theory and Techniques}, vol.~22, no.~1, pp. 48--52,  1974, doi:
  \href{http://dx.doi.org/10.1109/TMTT.1974.1128160}{10.1109/TMTT.1974.1128160}.

\bibitem{Banerjee2004}
S.~Banerjee and R.~Drayton, ``Circuit models for constant impedance
  micromachined lines on dielectric transitions,'' \emph{IEEE Transactions on
  Microwave Theory and Techniques}, vol.~52, no.~1, pp. 105--111,  2004, doi:
  \href{http://dx.doi.org/10.1109/TMTT.2003.921253}{10.1109/TMTT.2003.921253}.

\bibitem{Liu2016}
S.~Liu \emph{et~al.}, ``New methods for series-resistor calibrations on
  substrates with losses up to 110 ghz,'' \emph{IEEE Transactions on Microwave
  Theory and Techniques}, vol.~64, no.~12, pp. 4287--4297,  2016, doi:
  \href{http://dx.doi.org/10.1109/TMTT.2016.2609911}{10.1109/TMTT.2016.2609911}.

\bibitem{Amakawa2019}
S.~Amakawa \emph{et~al.}, ``Causal characteristic impedance determination using
  calibration comparison and propagation constant,'' in \emph{2019 92nd ARFTG
  Microwave Measurement Conference (ARFTG)},  2019, doi:
  \href{http://dx.doi.org/10.1109/ARFTG.2019.8637225}{10.1109/ARFTG.2019.8637225}.
  pp. 1--6.

\bibitem{IEC2020}
IEC, ``Measurement of the complex permittivity for low-loss dielectric
  substrates balanced-type circular disk resonator method,'' TC 46/SC 46F - RF
  and microwave passive components, International Standard IEC 63185:2020,
  2020.

\bibitem{BakerJarvis2005}
J.~Baker-Jarvis, M.~Janezic, B.~Riddle, R.~Johnk, C.~Holloway, R.~Geyer, and
  C.~Grosvenor, \emph{\BIBforeignlanguage{en}{Measuring the Permittivity and
  Permeability of Lossy Materials: Solids, Liquids, Metals, and negative-Index
  Materials}}, National Institute of Standards and Technology, Gaithersburg,
  MD, Feb. 2005.

\bibitem{Kato2019}
Y.~Kato and M.~Horibe, ``Broadband permittivity measurements up to 170-ghz
  using balanced-type circular-disk resonator excited by 0.8-mm coaxial line,''
  \emph{IEEE Transactions on Instrumentation and Measurement}, vol.~68, no.~6,
  pp. 1796--1805,  jun 2019, doi:
  \href{http://dx.doi.org/10.1109/TIM.2018.2886864}{10.1109/TIM.2018.2886864}.

\bibitem{GuidesinMetrology2011}
\BIBentryALTinterwordspacing
\emph{JCGM 102: Evaluation of Measurement Data - Supplement 2 to the "Guide to
  the Expression of Uncertainty in Measurement" - Extension to Any Number of
  Output Quantities}, Joint Committee for Guides in Metrology (JCGM), 2011.
  [Online]. Available:
  \url{http://www.bipm.org/utils/common/documents/jcgm/JCGM_102_2011_E.pdf}
\BIBentrySTDinterwordspacing

\bibitem{Hatab2022a}
Z.~Hatab, A.~B.~A. Alterkawi, H.~Takahashi, M.~Gadringer, and W.~Bösch,
  ``Low-return loss design of pcb probe-to-microstrip transition for
  frequencies up to 150 ghz,'' in \emph{2022 Asia-Pacific Microwave Conference
  (APMC)},  2022, doi:
  \href{http://dx.doi.org/10.23919/APMC55665.2022.9999913}{10.23919/APMC55665.2022.9999913}.
  pp. 208--210.

\bibitem{Megtron}
\BIBentryALTinterwordspacing
\emph{Megtron 7 IPC Standards R-5785(N) Data Sheet}, Panasonic, Apr. 2022.
  [Online]. Available:
  \url{https://industrial.panasonic.com/content/data/EM/PDF/ipcdatasheet_R-5785(N)_new.pdf}
\BIBentrySTDinterwordspacing

\bibitem{Scheer2020}
\BIBentryALTinterwordspacing
F.~Scheer, ``Advanced pcb material development for 5g and mm wave
  applications,'' Royal Circuits Webinar, Jun. 2020. [Online]. Available:
  \url{https://www.royalcircuits.com/2020/06/18/high-speed-materials/}
\BIBentrySTDinterwordspacing

\bibitem{Shiffler1980}
R.~E. Shiffler and P.~D. Harsha, ``Upper and lower bounds for the sample
  standard deviation,'' \emph{Teaching Statistics}, vol.~2, no.~3, pp. 84--86,
  1980, doi:
  \href{http://dx.doi.org/10.1111/j.1467-9639.1980.tb00398.x}{10.1111/j.1467-9639.1980.tb00398.x}.

\bibitem{Arsenovic2022}
A.~Arsenovic \emph{et~al.}, ``scikit-rf: An open source python package for
  microwave network creation, analysis, and calibration [speaker’s corner],''
  \emph{IEEE Microwave Magazine}, vol.~23, no.~1, pp. 98--105,  2022, doi:
  \href{http://dx.doi.org/10.1109/MMM.2021.3117139}{10.1109/MMM.2021.3117139}.

\bibitem{Hatab2022}
Z.~Hatab, M.~Gadringer, and W.~Bösch, ``Improving the reliability of the
  multiline trl calibration algorithm,'' in \emph{2022 98th ARFTG Microwave
  Measurement Conference (ARFTG)},  2022, doi:
  \href{http://dx.doi.org/10.1109/ARFTG52954.2022.9844064}{10.1109/ARFTG52954.2022.9844064}.
  pp. 1--5.

\bibitem{Hatab2022b}
------, ``Propagation of measurement and model uncertainties through multiline
  trl calibration,'' in \emph{2022 Conference on Precision Electromagnetic
  Measurements (CPEM)}, 2022, pp. 1--2.

\bibitem{Hatab2023}
Z.~Hatab, M.~E. Gadringer, and W.~Bösch, ``Propagation of linear uncertainties
  through multiline thru-reflect-line calibration,'' \emph{IEEE Transactions on
  Instrumentation and Measurement}, vol.~72, pp. 1--9,  2023, doi:
  \href{http://dx.doi.org/10.1109/TIM.2023.3296123}{10.1109/TIM.2023.3296123}.

\bibitem{Vardapetyan2008}
L.~Vardapetyan, J.~Manges, and Z.~Cendes, ``Sensitivity analysis of
  s-parameters including port variations using the transfinite element
  method,'' in \emph{2008 IEEE MTT-S International Microwave Symposium Digest},
   2008, doi:
  \href{http://dx.doi.org/10.1109/MWSYM.2008.4633219}{10.1109/MWSYM.2008.4633219}.
  pp. 527--530.

\bibitem{Savitzky1964}
A.~Savitzky and M.~J.~E. Golay, ``Smoothing and differentiation of data by
  simplified least squares procedures.'' \emph{Analytical Chemistry}, vol.~36,
  no.~8, pp. 1627--1639,  1964, doi:
  \href{http://dx.doi.org/10.1021/ac60214a047}{10.1021/ac60214a047}.

\end{thebibliography}
